\documentclass[aps, prd, reprint, twocolumn, nofootinbib, superscriptaddress]{revtex4-1}

\usepackage{hyperref}
\usepackage{amsmath,amssymb}
\usepackage{graphicx,psfrag,float,epsfig,color}
\usepackage{epstopdf}
\usepackage{comment}
\usepackage[usenames,dvipsnames,svgnames,table]{xcolor}
\usepackage{multirow}

\DeclareMathOperator{\sgn}{sgn}

\begin{document}


\title{Self-force via Green functions and worldline integration}

\author{Barry Wardell}
\affiliation{School of Mathematical Sciences and Complex \& Adaptive Systems Laboratory, University College Dublin, Belfield, Dublin 4, Ireland} 
\affiliation{Department of Astronomy, Cornell University, Ithaca, NY 14853, USA}

\author{Chad R.\!~Galley}
\affiliation{Theoretical Astrophysics, California Institute of Technology, Pasadena, California USA} 

\author{An{\i}l Zengino\u{g}lu}
\affiliation{Theoretical Astrophysics, California Institute of Technology, Pasadena, California USA} 

\author{Marc Casals}
\affiliation{Department of Cosmology, Relativity and Astrophysics (ICRA), Centro Brasileiro de Pesquisas F\'isicas,  
Rio de Janeiro, CEP 22290-180, Brazil.} 

\author{Sam R.\!~Dolan}
\affiliation{Consortium for Fundamental Physics, School of Mathematics and Statistics,
University of Sheffield, Hicks Building, Hounsfield Road, Sheffield S3 7RH, United Kingdom.}	

\author{Adrian C.\!~Ottewill}
\affiliation{School of Mathematical Sciences and Complex \& Adaptive Systems Laboratory, University College Dublin, Belfield, Dublin 4, Ireland} 

\begin{abstract}
A compact object moving in curved spacetime interacts with its own gravitational field. 
This leads to both dissipative and conservative
 corrections to the motion, which can be interpreted as
a self-force acting on the object. 
The original formalism 
describing this self-force relied heavily on the Green function of the linear differential operator that 
governs gravitational perturbations. However, because the global calculation of Green functions in non-trivial 
black hole spacetimes has been an open problem until recently, alternative methods were established 
to calculate self-force effects
using sophisticated regularization techniques that avoid the computation of the global Green function.
We present a method for calculating the self-force that employs the global Green function and is 
therefore closely modeled after the original self-force expressions.
Our quantitative method involves two stages: (i) numerical approximation of the retarded Green 
function in the background spacetime; (ii) evaluation of convolution integrals along the worldline of the 
object. 
This novel approach can be used along arbitrary worldlines, including those currently inaccessible to 
more established computational techniques. Furthermore, it yields geometrical insight into the contributions 
to self-interaction from curved geometry (back-scattering) and trapping of null geodesics.
We demonstrate the method on the motion of a scalar charge in Schwarzschild spacetime. 
This toy model retains the physical history-dependence of the self-force but avoids gauge 
issues and allows us to focus on basic principles.
We compute the self-field and self-force for many worldlines 
including accelerated circular orbits, eccentric orbits at the separatrix, and radial infall. 
This method, closely modeled after the original formalism, provides a promising complementary 
approach to the self-force problem. 
\end{abstract}

\maketitle

\section{Introduction}
\label{sec:intro}

Gravitational waves emitted by binary systems featuring black holes contain a wealth of information about gravity in the strong field regime. Of particular interest is the case of a compact body of mass $m$ (e.g.~a neutron star or black hole) in orbit around a larger black hole of mass $M$, such that $m / M \ll 1$. In this scenario, the radiation reaction timescale is much longer than the orbital period, and the system undergoes many cycles in the vicinity of the innermost stable circular orbit (typically, $\sim M/m$ in the final year before coalescence). The gravitational wave signal provides a direct probe of the spacetime of the massive black hole.

In the self-force interpretation, the compact body's motion is associated with a worldline defined on the background spacetime of the larger partner. In the test-body limit $m = 0$, that worldline is a geodesic of the background. For a finite mass, the worldline is accelerated by a \emph{gravitational self-force} as the compact body interacts with its own gravitational field. The self-force has a dissipative and a conservative part, which drive an inspiraling worldline.

The calculation of the self-force in gravitational physics is an important problem from both a
fundamental and an astrophysical point of view (see reviews \cite{Barack:2009ux, Poisson:2011nh}).
Fundamentally, knowing the gravitational self-force gives insight into the basic physical
processes that determine the motion of a compact object in a curved background spacetime. This
problem has its roots in the motion of an electrically-charged particle in flat spacetime \cite{Dirac}, though
the physics of the gravitational self-force and electromagnetic radiation reaction are quite
different. Astrophysically, the gravitational self-force is important because it drives the
inspiral of a binary with an extremely small mass ratio
($\sim 10^{-5}$--$10^{-7}$). Such systems are expected gravitational wave sources for space-based
detectors (e.g., eLISA \cite{AmaroSeoane:2012km, AmaroSeoane:2012je}) that will provide detailed
information about the evolution of galactic mergers, general relativity in the strong-field regime,
and possibly the nature of dark energy, among others \cite{yellowbook}.

An expression for the gravitational self-force was formulated first in 1997 by Mino, Sasaki \&
Tanaka \cite{Mino:1996nk} and Quinn \& Wald
\cite{Quinn:1996am}. Working independently, they obtained the MiSaTaQuWa (pronounced
Mee-sah-tah-kwa) equation: an expression for the self-force at first-order in the mass ratio (in recent years, a more rigorous foundation for this formula has been established
\cite{Gralla:2008fg, Pound:2009sm, Pound:2010pj} and second-order extensions in $m/M$ have been
proposed \cite{Detweiler:2011tt, Pound:2012nt, Gralla:2012db, Pound:2012dk}). The MiSaTaQuWa force
can be split into ``instantaneous'' and ``history-dependent'' terms,
\begin{align}
	F^{\alpha} = F^{\alpha}_{\text{inst}} + F^{\alpha}_{\text{hist}}.
\end{align}
The instantaneous terms encapsulate the interaction between the source and the local spacetime geometry, 
and appear at a higher than leading order in the mass ratio.
The history-dependent term involves the worldline integration of the gradient of the retarded
Green function, which indicates a dependence on the past history of the source's motion.
The integral extends to the infinite past and is truncated just before coincidence at $\tau - \tau' =
\epsilon \rightarrow 0^+$. The dependence on the history of the source is due to gravitational
perturbations that were emitted in the past by the source.

Describing and understanding gravitational self-force effects is complicated due to the gauge freedom and
the computational burdens from the tensorial structure of gravitational perturbations. An important
stepping stone for computing gravitational self-force has historically been scalar models. These
models avoid the technical complications of the gravitational problem but still capture some
essential features of self-force, in particular, the history-dependence.

The retarded Green function (RGF) for the scalar field is defined as the fundamental solution $G_{\rm ret} (x, x')$
which yields causal solutions of the inhomogeneous scalar wave equation
\begin{align}
\Box_x  \Phi(x) = -4\pi \rho (x)
\end{align}
through
\begin{align}
\Phi(x) =\int G_{\rm ret} (x, x') \rho(x') \sqrt{-g(x')} \, d^4 x' .
\end{align}
Here, $G_{\rm ret} (x, x') $ is a distribution which acts on test functions $\rho(x')$ and also
depends on the field point $x$, with the property that its support is $x'\in J^-(x)$, where $J^-(x)$ is the 
causal past of $x$. This can be expressed as
\begin{equation}
\label{eq:green}
\Box_x G_{\rm ret}(x, x') = - 4 \pi \delta_4(x, x')\,,
\end{equation}
along with appropriate boundary conditions. Here,
$\delta_4(x,x') = \delta_4(x-x') / \sqrt{-g}$ is the
four-dimensional invariant Dirac delta distribution and $g$ is the
determinant of the metric $g_{\mu\nu}$. 

It was shown by Quinn \cite{Quinn:2000wa} that the self-field of a particle with scalar charge
$q$ depends on the past history of the particle's motion through
\begin{align}
	\Phi_{\rm hist} (z^\lambda) = {} & q \lim_{\epsilon \to 0^+} \int_{-\infty}^{\tau- \epsilon} G_{\rm ret} (z^\lambda, z^{\lambda'} ) \, d\tau' \, .
\label{eq:Phisym1}
\end{align}
Here, $z^\lambda(\tau)$ describes the particle's worldline parameterized by proper time $\tau$
 and a prime on an index means that it is associated with the
worldline parameter value $\tau'$ (e.g., $z^{\alpha'} = z^\alpha (\tau')$).
Likewise, Quinn showed that the self-force experienced by the particle
can be written as $F^\alpha = F^\alpha_{\text{inst}} + F^\alpha_{\text{hist}} $
where $F^\alpha_{\text{inst}}$ is the ``instantaneous'' force, which vanishes for geodesic motion in a
vacuum spacetime. The history-dependent term is
\begin{equation}
F^\alpha_{\text{hist}} (z^\lambda)  = q^2 g^{\alpha\beta} \lim_{\epsilon \rightarrow 0^+} \int_{-\infty}^{\tau -\epsilon} \nabla_\beta G_{\rm ret}(z^\lambda, z^{\lambda'}) \,d\tau' .
\label{eq:Fsym1}
\end{equation}
The covariant derivative is taken with respect to the first argument of the RGF. 
The main difficulty 
in computing the self-field and the self-force through \eqref{eq:Phisym1} and \eqref{eq:Fsym1}, respectively,
is the computation of the 
RGF along the entire past worldline of the particle.

In addition to its role in computing the self-force the Green function also plays an
essential part in understanding wave propagation.
When the Green function to a linear partial differential operator,
such as the wave operator, is known, 
any concrete solution with arbitrary initial data and source can be constructed 
via a simple convolution. Naturally, much effort has
been devoted to the study and the calculation of Green functions in curved
spacetimes, but a sufficiently accurate, quantitative description of its global behavior has remained
an open problem until recently.

There are various difficulties regarding the calculation of Green
functions in curved spacetimes. In four dimensional flat spacetime
the RGF is supported \emph{on} the lightcone
only (when the field is massless), 
whereas in curved spacetimes the RGF has extended support also
\emph{within} the lightcone (here ``lightcone'' refers to the set of points
$x$ connected to $x'$ via null geodesics). Further, the lightcone
generically self-intersects along \emph{caustics} due to focusing
caused by spacetime curvature.

These rich features of the RGF make it difficult to compute globally. A local approximation
is not sufficient to evaluate the integral in \eqref{eq:Fsym1} because the amplitude at a point on the 
worldline depends on the entire past history of the source. Early attempts 
\cite{Anderson:2003qa,Anderson:2004eg,Anderson:2005ds,Ottewill:2007mz,Ottewill:2008uu} 
were founded upon the Hadamard parametrix \cite{Hadamard}, which gives the RGF in the form 
\begin{align} 
	\label{eq:Hadamard}
	G_{\rm ret}(x,x') = \Theta_-(x,x') \big[ U(x,x') \delta(\sigma) - V(x,x') \Theta(-\sigma) \big] . 
\end{align}
Here $U$ and
$V$ are smooth biscalars, $\Theta(\cdot)$ is the Heaviside function,
$\Theta_-(x,x')$ is unity when $x'$ is in the past of $x$ (and
zero otherwise), and $\sigma=\sigma(x,x')$ is the Synge world-function
(i.e., one-half of the squared distance along the geodesic connecting $x$ and
$x'$). The Hadamard
parametrix is only valid in the region in which $x$ and $x'$
are connected by a unique geodesic (more precisely, if $x$ lies within
a causal domain, commonly referred to using the less precise term \emph{normal neighborhood}, of $x'$ \cite{Friedlander:1975}). 
In strongly curved spacetimes the contribution from outside the normal neighbourhood cannot be
neglected. Therefore, the Hadamard parametrix is insufficient for
computing worldline convolutions.\footnote{See~\cite{Casals:2012px} for a proposal to calculate the global RGF using
convolutions of the Hadamard parametrix, although  difficult
for practical applications in black hole spacetimes.} A method of matched expansions 
was outlined \cite{Poisson:Wiseman:1998,Anderson:2005gb}, in which a
quasi-local expansion for $G_{\rm ret}$, based on the Hadamard parametrix, would
be matched onto a complementary expansion valid in the more distant
past.

Despite initial promise, this idea for the global evaluation of the
RGF proved difficult to implement, and other schemes for computing self-force
came to prominence: the mode-sum regularization method \cite{Barack:1999wf} 
and the effective source method \cite{Vega:2007mc,Barack-Golbourn}.
A practical advantage of these methods is that they work at the
level of the (sourced) field rather than requiring global knowledge of
RGFs. These methods require a regularization procedure based on
a local analysis of the Green function by a decomposition into the
singular and regular fields given by Detweiler \& Whiting
\cite{Detweiler:2002mi}. They have yielded impressive results
\cite{Burko:1999zy, Burko:2000xx, Burko:2000yx, Barack:2000zq, Detweiler:2002gi, DiazRivera:2004ik, Vega:2009qb, Haas:2007kz, Burko:2001kr, Warburton:2010eq, Warburton:2011hp, Keidl:2006wk, Barack:2002ku, Barack:2007tm, Detweiler:2008ft, Barack:2009ey, Shah:2010bi, Canizares:2010yx, Keidl:2010pm, Dolan:2010mt, Haas:2011np, Dolan:2011dx, Barack:2011ed, Shah:2012gu, Dolan:2012jg, Dolan:2013roa},
including the computation of self-consistent orbits at first-order in the mass ratio
\cite{Warburton:2011fk, Diener:2011cc, Lackeos:2012de}. 
See \cite{Barack:2009ux} for a recent review of numerical self-force computations.

The steady progress in the computation of the self-force compared with the lack of
quantitative results on global Green functions may have
led to the notion that the computation of the self-force via the integration of the 
RGF was not practicable. Nevertheless, there has been recent work
toward the global calculation of the RGF. The method of matched expansions for
history-integral evaluation was revisited in \cite{Casals:2009zh}, where
it was shown that a distant-past expansion for the GF could be
used to compute the self-force in the Nariai spacetime (a black hole toy model). 
Subsequently, steps were taken towards overcoming the technical difficulties with applying the
analysis on realistic black-hole spacetimes
\cite{Dolan:2011fh,Casals:2012ng,Casals:2012tb}.

An interesting by-product of these investigations has been an improved understanding of the
structure of the RGF in spacetimes with caustics. While it was already known
(e.g.,~\cite{Garabedian,Ikawa}) that singularities in the global RGF occur when the two spacetime
points are connected via a null geodesic, the specific form of the singularities beyond the normal
neighborhood was not previously known within general relativity.
Ori \cite{Ori} noted
that the singular part of the RGF undergoes a transition each time the wavefront encounters a
caustic, typically cycling through a four-fold sequence $\delta(\sigma) \rightarrow 1/(\pi \sigma)
\rightarrow -\delta(\sigma) \rightarrow -1/(\pi \sigma) \rightarrow \delta(\sigma)$ (there are some
exceptions to this four-fold cycle). 
Such four-fold structure can be understood as the  wavefront picking up a phase $-\pi/2$ every time
it crosses a caustic~\cite{Casals:2009zh}. In fact, this property has first been discovered in optics 
in the late 19th century \cite{Gouy} and is related to the so-called Maslov 
index~\cite{Maslov'65,MaslovWKB}. The phase transitions were confirmed via analytic methods on
Schwarzschild spacetime \cite{Dolan:2011fh} and black hole toy model space
times~\cite{Casals:2009zh,Casals:2012px}. A general mathematical analysis of the effect of caustics
on wave propagation in general relativity was given in \cite{Harte:2012uw}. 

Two recent developments along this line of research play an essential role for this paper. 
The first one is the numerical approximation of the global Green function in Schwarzschild spacetime 
\cite{Zenginoglu:2012xe}. By replacing the delta-distribution source in \eqref{eq:green} by a narrow Gaussian and evolving the scalar wave equation numerically, it was possible to provide the first global approximation of the RGF \cite{Zenginoglu:2012xe}. The source in \eqref{eq:green} generates
``echoes'' of itself due to trapping at the photon sphere, which were called ``caustic echoes'' in \cite{Zenginoglu:2012xe}. An ideal observer at null infinity first measures the direct signal from the delta-distribution source, and later encounters, at nearly regular time intervals, exponentially-decaying caustic echoes. Both the arrival intervals and the exponential decay 
are determined by the properties of null geodesics around the photon sphere (their travel time and Lyapunov exponent).
This work also provided a physical understanding of the four-fold sequence in caustic echoes:
the trapping causes the wavefront to intersect itself at caustics 
as encapsulated by a Hilbert transform \cite{Zenginoglu:2012xe}, which is equivalent to the $-\pi/2$ 
phase shift observed previously. 
A two-fold cycle was also observed and explained in \cite{Zenginoglu:2012xe,Casals:2012px} 
whenever the field point was in the equatorial plane (with the source) at an azimuthal angle of 
$\varphi = n \pi$ for $n$ an integer. Eventually, the echoes diminish below the late-time tail.

The second development is the culmination of previous semi-analytical efforts 
\cite{Casals:2009zh,Dolan:2011fh,Casals:2012ng,Casals:2012tb} in the calculation of the global RGF
in Schwarzschild spacetime using the method of matched expansions \cite{Casals:2013mpa}. 
The RGF was calculated semi-analytically in the more distant past using a Fourier-mode decomposition 
and contour-deformation in the complex frequency plane, which also allowed for the computation of the 
self-force on a scalar charge in Schwarzschild spacetime. 

In this paper, we combine a numerically computed global approximation to the RGF
based on \cite{Zenginoglu:2012xe} with semi-analytic approximations at early and late times 
\cite{Casals:2013mpa} to compute
the self-force for arbitrary motion via the worldline convolution integral, Eq.~\eqref{eq:Fsym1}.
To demonstrate the basic principles, we focus on the self-force 
on a scalar charge in a Schwarzschild background spacetime.

Our calculations consist of two parts. First, we construct quantitative global approximations of
the Green function in Schwarzschild spacetime by replacing the delta-distribution by a narrow Gaussian 
\cite{Zenginoglu:2012xe}. We augment the numerical calculation with analytical approximations through 
quasi-local expansions at early times (\cite{Ottewill:2007mz, Wardell:2009un, Casals:2009xa}) and
through branch-cut integrals at late times \cite{Casals:2012tb,Casals:Ottewill:2014}. Second, we directly 
evaluate the convolution integrals for the self-field \eqref{eq:Phisym1} and the self-force \eqref{eq:Fsym1} at the desired 
point along the given worldline. We discuss a variety of orbital configurations, including accelerated 
circular orbits, eccentric geodesic orbits on the separatrix, and radial infall. 
 
Beyond its direct relation to the original formalism, there are two additional motivations 
for computing the self-force through worldline convolutions of the Green function. 
The first motivation is conceptual: the method allows us to answer questions such as,
how does the self-force depend on the past-history, and how far back in the history must one
integrate to obtain an accurate estimate of the convolution. The second motivation is
complementarity: the method is well-suited to computing the self-force along {\it aperiodic} (or nearly aperiodic) trajectories, such as unbound orbits, highly-eccentric or zoom-whirl orbits, and ultra-relativistic
trajectories. These motions are difficult for or inaccessible to existing methods.

We believe this work challenges a perception that the original formulation by MiSaTaQuWa is
ill-suited to practical calculations. We aim to demonstrate that the new method has the
potential to make self-force calculations via worldline convolutions of Green functions as routine
and as accurate as (time-domain) computations via the more established methods of mode-sum
regularization and effective source.

In Sec.~\ref{sec:methods} we describe methods for constructing the RGF with numerical
(\ref{sec:numerical}) and semi-analytic (\ref{sec:analytical}) approaches. In
Sec.~\ref{sec:results} we validate our method (\ref{sec:validation}) and present a selection of
results for three examples of motion: accelerated circular orbits (\ref{sec:accelerated}),
eccentric geodesics (\ref{sec:eccentric}), and radial infall (\ref{sec:infall}). We conclude with a
discussion in Sec.~\ref{sec:conclusion}. Throughout we choose conventions with $G=c=1$ and metric
signature $(-+++)$.

\section{Construction of the RGF}
\label{sec:methods}

The most difficult technical step in our computation of self-force is the construction of the global
RGF. In Sec.~\ref{sec:numerical} we approximate the Green function numerically by replacing 
the delta-distribution by a narrow Gaussian in the initial data using the Kirchhoff formalism. We exploit the 
spherical symmetry of the background by performing a spherical harmonic decomposition and solving 
(1+1) dimensional wave equations for the Green function using standard numerical methods along with
double hyperboloidal layers. 
In Sec.~\ref{sec:analytical} we augment the numerical 
solution at early and late times with semi-analytic approximations based on the quasi-local
expansion and late-time behavior.

\subsection{Numerical computation of the RGF}
\label{sec:numerical}

The global construction of the RGF in \cite{Zenginoglu:2012xe} used a narrow Gaussian to approximate 
the 4-dimensional Dirac delta-distribution source in 
\eqref{eq:green} as, 
\begin{align}
	\delta_4 (x-x') \approx  \frac{ 1}{ (2\pi \bar{\varepsilon}^2)^2 } \exp \! \left[ - \sum_{\alpha=0}^3  \frac{ (x^\alpha-x'{}^\alpha)^2 }{ 2\bar{\varepsilon}^2 } \right].
\label{eq:gaussian1}
\end{align}
Here, $\bar{\varepsilon}$ is the width of the Gaussian centered at the base point $x'$. 
The numerical construction in  \cite{Zenginoglu:2012xe} is performed in (3+1) dimensions and 
does not rely on (or exploit) any symmetries. 
In this paper we use the Gaussian approximation not in the source but in initial data. 
We use a Kirchhoff representation of the solution of the initial value 
problem corresponding to a given Cauchy surface in terms of the RGF
and then set Gaussian initial data to recover an approximation to the RGF.
The Kirchhoff representation can also be used in (3+1)-dimensions. However, in a subsequent step 
we perform an $\ell$-mode decomposition to exploit the spherical symmetry of the background, and 
reduce the wave equation to (1+1) dimensions, rather than (3+1) dimensions. 

To summarize, our approach differs from \cite{Zenginoglu:2012xe} in the dimensionality of the time-domain wave equation 
[(1+1)- versus (3+1)-dimensions] and in the role of the approximated delta-distribution: 
as a source to the wave equation in \cite{Zenginoglu:2012xe} and as initial data in our approach.

\subsubsection{RGF from impulsive initial data}
\label{sec:Kirchhoff}

Given Cauchy data on a spatial hypersurface $\Sigma$, the Kirchhoff representation in terms of the
RGF can be used to determine the solution of the homogeneous wave equation,\footnote{
We use $\Psi$ to denote the field associated with our numerical approximation of the RGF to distinguish it 
from the approximated inhomogeneous solution to the wave equation, 
$\Phi$, that results from convolving $\Psi$ with a generic source.}
\begin{equation}
\label{eq:homogeneous-wave}
  \Box \Psi = 0
\end{equation}
at an arbitrary point $x'$ in the past of $\Sigma$,
\begin{align}
\label{eq:Kirchhoff}
\Psi(x') = {} &  -\frac{1}{4\pi} \int_\Sigma \big[ G_{\rm ret}(x, x') \nabla^\mu \Psi(x) \nonumber \\
	& {\hskip0.775in} - \Psi(x) \nabla^\mu G_{\rm ret}(x, x') \big] d \Sigma_\mu.
\end{align}
Here, $d\Sigma_\mu = n_\mu \sqrt{h}\, d^3 \mathbf{x}$ is the future-directed surface element on $\Sigma$, where $n_\mu=-\alpha \delta_{\mu}^0$ is the future-directed unit normal to the surface, $\alpha$ is the
lapse, and $h$ is the determinant of the induced metric. Using reciprocity, there is also an
equivalent representation in terms of the advanced Green function but we are interested 
in the RGF in this paper.

We choose a time coordinate $t$ (not necessarily Schwarzschild time) so that the
surface $\Sigma$ corresponds to $t=t_0$. Setting initial data
\begin{subequations}
\label{eq:initial-data} 
\begin{align}
  &\Psi(x)|_\Sigma = \Psi(t_0,\mathbf{x})= 0\\
  &n_\mu \nabla^\mu \Psi(x)|_\Sigma = n_\mu \nabla^\mu \Psi(t_0,\mathbf{x})= -4\pi \delta_3\bigl(\mathbf{x},\mathbf{x}_0\bigr)\,,
\end{align}
\end{subequations}
Eq.~\eqref{eq:Kirchhoff} reduces to
\begin{align}
\Psi(x')
 &= \int_\Sigma G_{\rm ret}(t_0,\mathbf{x};x') \delta_3(\mathbf{x},\mathbf{x}_0) \sqrt{h} \, d^3 \mathbf{x} \nonumber \\
 &= G_{\rm ret}(t_0,\mathbf{x}_0; x').
\end{align}
In other words, by evolving the homogeneous wave equation backwards in time with appropriate
impulsive initial data, we obtain the RGF with the point $(t_0,\mathbf{x}_0)$
fixed on the hypersurface $\Sigma$ for all values of $x'$ to the past. Thus, with just one
calculation we obtain the RGF for fixed point $(t_0,\mathbf{x}_0)$ and 
\emph{all possible} source points $z(\tau')$ for trajectories
passing through $(t_0,\mathbf{x}_0)$ appearing in the self-force history-integral.
Note that we have been careful to describe this as fundamentally evolving \emph{backwards} in time.
The alternative---evolving Cauchy data \emph{forwards} in time---would have instead yielded
$G_{\rm ret}(x'; t_0,\mathbf{x}_0)$, which would be impractical for worldline convolutions as it
would require a separate simulation for each point in the convolution integral. For
time-reversal-invariant backgrounds such as the Schwarzschild spacetime the two procedures
are, in fact, equivalent, but to emphasize the general applicability of the method we do not exploit
that property here.

We can also compute derivatives of the Green function with
respect to $x'$ by simply differentiating the evolved field $\Psi(x')$. However, in order to compute
derivatives with respect to the source point $x$, as is demanded by a self-force calculation, it 
is necessary to modify the scheme by setting
\begin{subequations}
\label{eq:initial-data-derivative}
\begin{align}
  &\Psi(x)|_\Sigma = \Psi(t_0,\mathbf{x})= 0\\
  &n_\mu \nabla^\mu \Psi(x)|_\Sigma = n_\mu \nabla^\mu \Psi(t_0,\mathbf{x})= 4\pi \partial_i  \delta_3\bigl(\mathbf{x},\mathbf{x}_0\bigr)\,.
\end{align}
\end{subequations}
Integrating by parts, we have
\begin{align}
\Psi(x')
&= \int_\Sigma \partial_i  G_{\rm ret}(t_0,\mathbf{x}; x') \delta_3(\mathbf{x},\mathbf{x}_0) \sqrt{h} \, d^3 \mathbf{x} \nonumber \\
  &=\partial_i G_{\rm ret}(t_0,\mathbf{x}_0; x')\,,
\end{align}
and we obtain the spatial derivatives of the Green function, as required. Finally, for the time
derivative we set
\begin{subequations}
\label{eq:initial-data-time-derivative}
 \begin{align}
  &\Psi(x)|_\Sigma = \Psi(t_0,\mathbf{x}) =4\pi \delta_3\bigl(\mathbf{x},\mathbf{x}_0\bigr)\\
 &n_\mu \nabla^\mu \Psi(x)|_\Sigma = n_\mu \nabla^\mu \Psi(t_0,\mathbf{x})=  0\,,
\end{align}
\end{subequations}
to obtain
\begin{align}
\Psi(x')
 &= \int_\Sigma n_\mu \nabla^\mu G_{\rm ret}(t_0,\mathbf{x};x') \delta_3(\mathbf{x},\mathbf{x}_0)\sqrt{h} \, d^3 \mathbf{x}  \nonumber \\
  &=n_\mu \nabla^\mu G_{\rm ret}(t_0,\mathbf{x}_0;x')\,.
\end{align}

\subsubsection{Decomposition for 1+1 dimensions}

In spherically symmetric spacetimes, we exploit the spherical symmetry of the background to
develop an efficient scheme for solving \eqref{eq:homogeneous-wave} with initial data given by 
\eqref{eq:initial-data}, \eqref{eq:initial-data-derivative}, and
\eqref{eq:initial-data-time-derivative}. By decomposing the Green function into spherical harmonics
and applying the Kirchhoff formula to the part depending on time and the radius, we
significantly improve the efficiency of our calculations. 

Consider the decomposition of the field into spherical harmonic $(\ell, m)$ modes,
\begin{equation}
  \Psi(t,r,\theta,\varphi) =
    \sum_{\ell=0}^\infty \sum_{m=-\ell}^{\ell} \frac{1}{r} \Psi_{\ell m}(t,r) Y_{\ell m} (\vartheta, \varphi).
\end{equation}
The spherical symmetry of the problem allows us to place the base point $x_0$ on the axis, in which
case only the $m=0$ modes contribute. Then the spherical harmonic decomposition reduces to
\begin{equation}
\label{eq:field-Legendre}
  \Psi(t,r,\gamma) =
    \sum_{\ell=0}^\infty \frac{1}{r} (2\ell+1)\Psi_{\ell}(t,r) P_{\ell} (\cos\gamma),
\end{equation}
where $\Psi_{\ell}\equiv \Psi_{\ell m=0}$. Likewise, we decompose the Green function and its derivative by
\begin{equation}
  \label{eq:G_sum}
  G_{\rm ret}(t_0,\mathbf{x}_0;x') =
     \frac{1}{r_0 r'} \sum_{\ell=0}^\infty (2\ell+1) G_{\ell}(t_0, r_0; t',r') P_{\ell} (\cos\gamma),
\end{equation}
and
\begin{align}
  \label{eq:dGdr_sum}
  \partial_r &G(t_0,\mathbf{x}_0;x') = \nonumber \\
    & \frac{1}{r'} \sum_{\ell=0}^\infty (2\ell+1) P_{\ell} (\cos\gamma) \left\{\partial_r \left[\frac{1}{r} G_{\ell}(t_0, r; t',r')\right]\right\}_{r=r_0}.
\end{align}

Choosing the standard Schwarzschild time $t$ and the tortoise coordinate
$r^\ast \equiv r+2M \ln (r/2M -1)$, the Kirchhoff formula 
\eqref{eq:Kirchhoff} reduces to
\begin{equation}
  \Psi_\ell(t',r') = \int_\Sigma G_\ell(t,r;t',r') \partial_t \Psi_\ell(t,r) dr^\ast.
\end{equation}
\begin{align}
  \Psi_\ell(t',r') = {}& \int_{-\infty}^{\infty} \big[ G_\ell(t,r;t',r')  \partial_t \Psi_\ell(t,r) \nonumber \\
  	& {\hskip0.775in} - \Psi_\ell(t,r) \partial_t G_\ell(t,r;t',r')  \big] dr^\ast
\end{align}
Setting initial data 
\begin{equation}
  \label{eq:initial-data-l}
 \Psi_\ell(t,r)|_\Sigma = 0, \quad \partial_t \Psi_\ell(t,r)|_\Sigma = \delta(r^\ast-r^\ast_0),
\end{equation}
yields
\begin{align}
\Psi_\ell(t',r') &= \int_{-\infty}^{\infty} G_\ell(t_0,r;t',r') \delta(r^\ast-r^\ast_0) dr^\ast \nonumber \\
  & = G_\ell(t_0, r_0; t', r').
  \label{eq:Phi=G}
\end{align}
Thus, by evolving the homogeneous $1$+$1$ dimensional wave equation
(see Eqs.~(\ref{eq:mol1}) and (\ref{eq:mol2}) below)
with this initial data, we obtain the $\ell$
modes of the retarded Green function with the retarded point $r_0$ fixed on the initial
hypersurface. Similarly, we can compute the radial derivative of the Green function by choosing
initial data
\begin{equation}
  \label{eq:initial-data-l-dr}
  \Psi_\ell(t,r)|_\Sigma =0, \quad \partial_t \Psi_\ell(t,r)|_\Sigma = - \frac{f}{r} \partial_r \left[f^{-1} \delta(r^\ast-r^\ast_0)\right]
\end{equation}
with $f \equiv 1-2M/r$. This gives
\begin{align}
\Psi_\ell(t',r') &= -\int_{-\infty}^{\infty} \frac{f}{r} G_\ell(t_0,r;t',r') \partial_r \left[f^{-1}\delta(r^\ast-r^\ast_0)\right] dr^\ast \nonumber \\
  & = \partial_r \left[\frac{1}{r}G_\ell(t_0, r; t', r')\right]_{r=r_0}.
  \label{eq:dr Phi=G}
\end{align}
Finally, to compute the $t$ derivative we set initial data
\begin{equation}
  \label{eq:initial-data-l-dt}
 \Psi_\ell(t,r)|_\Sigma = \delta(r^\ast-r^\ast_0), \quad \partial_t \Psi_\ell(t,r)|_\Sigma = 0,
\end{equation}
which yields
\begin{align}
\Psi_\ell(t',r') &= \int_{-\infty}^{\infty} \left[\partial_t G_\ell(t,r;t',r')\right]_{t=t_0} \delta(r^\ast-r^\ast_0) dr^\ast \nonumber \\
  & = \left[\partial_t G_\ell(t, r_0; t', r')\right]_{t=t_0}.
  \label{eq:dt Phi=G}
\end{align}
We obtain $G_{\ell}$ and its $t$ and $r$ derivatives by solving the corresponding
initial value problem. The $\gamma$ derivatives are trivially given by using \eqref{eq:Phi=G} in
\eqref{eq:G_sum} and differentiating the Legendre polynomials with respect to $\gamma$.

\subsubsection{Gaussian approximation to the Dirac delta distribution and the smooth angular cutoff}
\label{sec:gaussiandelta}

We approximate the Dirac delta distribution in the initial data 
(see Eqs.~(\ref{eq:initial-data-l}), (\ref{eq:initial-data-l-dr}) and (\ref{eq:initial-data-l-dt}))
on our numerical grid by a Gaussian of
finite width, $\varepsilon$. In the limit of zero width, the delta distribution is recovered,
\begin{equation}
 \delta(r^\ast-r^\ast_0) = \lim_{\varepsilon \to 0^+} \frac{1}{(2\pi\varepsilon^2)^{1/2}} e^{-(r^\ast-r^\ast_0)^2/2\varepsilon^2}
\end{equation}
This replacement effectively limits the shortest length scales which can be represented by the
spherical harmonic expansion of the field---in this way there is a direct correspondence
between the use of a finite-width Gaussian and a finite number of $\ell$ modes. Higher $\ell$
modes oscillate more and more rapidly both in space and time. Once the scale of these
oscillations becomes comparable to the width of the Gaussian, any higher $\ell$ modes are not
faithfully resolved. The advantage of using a smooth Gaussian, however, is that this cutoff happens
in a smooth fashion.

We cut off the formally infinite sum over $\ell$ in \eqref{eq:G_sum} and \eqref{eq:dGdr_sum} at
some finite $\ell_{\rm max}$. A sharp cutoff in a spectral expansion yields highly oscillatory,
unphysical features in the result. The replacement of a delta distribution with a smooth Gaussian
in our numerical scheme mitigates this somewhat; for a fixed Gaussian width $\varepsilon$ there
exists a finite $\ell_{\rm max}$ which is sufficient to eliminate the unphysical oscillations (for
a detailed discussion of the relation between $\varepsilon$ and $\ell_{\rm max}$ see Sec.~IV A 3 of
\cite{Casals:2013mpa}). However, for very small Gaussian widths this $\ell_{\rm max}$ may be
unpractically large.

In this work, as previously in~\cite{Casals:2009zh,Casals:2013mpa}, we have
employed a smooth sum method which is very effective in eliminating the high-frequency oscillations
while maintaining the physically-relevant low $\ell$ contribution important for computing
the self-force. The basic idea to to replace the sum in \eqref{eq:field-Legendre} (or, 
equivalently, in \eqref{eq:G_sum} and \eqref{eq:dGdr_sum}) with a smooth cut off at large $\ell$ \cite{Hardy},
\begin{equation}
\label{eq:field-Legendre-smooth}
  \Psi(t,r,\gamma) =
    \sum_{\ell=0}^{\ell_{\rm max}} e^{-\ell^2/2\ell_{\rm cut}^2} \frac{1}{r} (2\ell+1)\Psi_{\ell}(t,r) P_{\ell} (\cos\gamma),
\end{equation}
where we empirically
choose $\ell_{\rm cut} = \ell_{\rm max}/5$. 
The introduction of such a smoothing factor can be related~\cite{Casals:2013mpa} to: i) the
``smearing'' in the angular coordinate of the distributional features of the Green function,
and  ii) the replacement of the $\delta_4$-source in (\ref{eq:green}) by a ``narrow'' Gaussian 
distribution.

\begin{figure}
 \includegraphics[width=\columnwidth]{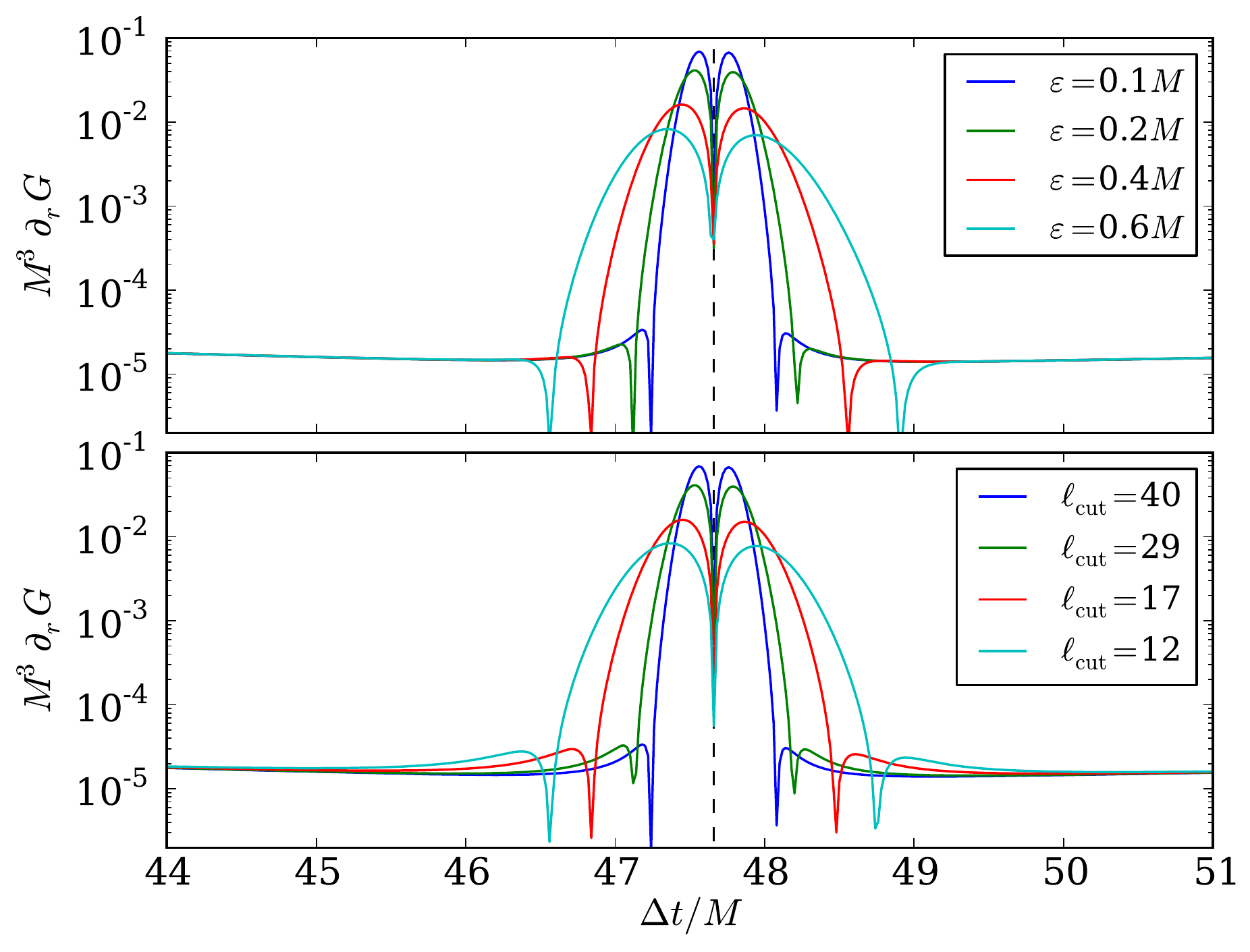}
 \caption{
Top: The radial derivative of the Green function for the $e=0.5$, $p=7.2$ eccentric orbit case
computed using Gaussians of varying widths and a fixed smooth-sum cutoff of $\ell_{\rm cut}=40$. As the width is decreased, the sharp features near the
null geodesic crossing around $\Delta t \approx 47.6M$ are better resolved. Away from the null geodesic crossings, the
improvement with decreasing width is much less significant.
Bottom: Since a finite width Gaussian also implies an effective maximum number of $\ell$ modes, a
very similar plot is obtained when the number of $\ell$ modes included is increased (with a fixed Gaussian width of $\varepsilon=0.1M$).
 }
 	\label{fig:gaussian}
\end{figure}

The effect of varying $\varepsilon$ and $\ell_{\rm cut}$ on the Green function can be seen in 
Fig.~\ref{fig:gaussian}. The sharp features near caustic echoes are resolved only for small $\varepsilon$ and
large $\ell_{\rm cut}$;  but away from caustic echoes the Green function is well resolved even for large 
$\varepsilon$ and small $\ell_{\rm cut}$.

Because the regularized self-field is a smooth function of spacetime, the spectral representation in terms 
of spherical-harmonic modes converges exponentially with the number of $\ell$ modes and the small-$\ell$
modes seem sufficient.  There are, however, some cases where this approximation is not very useful and
large $\ell$ features become important. For example, for ultra-relativistic orbits close to the Schwarzschild light ring at $r=3M$, the exponentially-convergent regime is deferred to very large $\ell$ \cite{Akcay:2012ea}. In such cases, it is likely that alternative methods which capture the large-$\ell$ structure 
are more suitable, including asymptotic expansions of special functions \cite{Casals:2012px, Nolan-Capra2013},
WKB methods \cite{Dolan:2011fh}, the geometrical optics approximation \cite{Zenginoglu:2012xe},
and frozen Gaussian beams \cite{LuYang}.

\begin{figure}
 \includegraphics[width=\columnwidth]{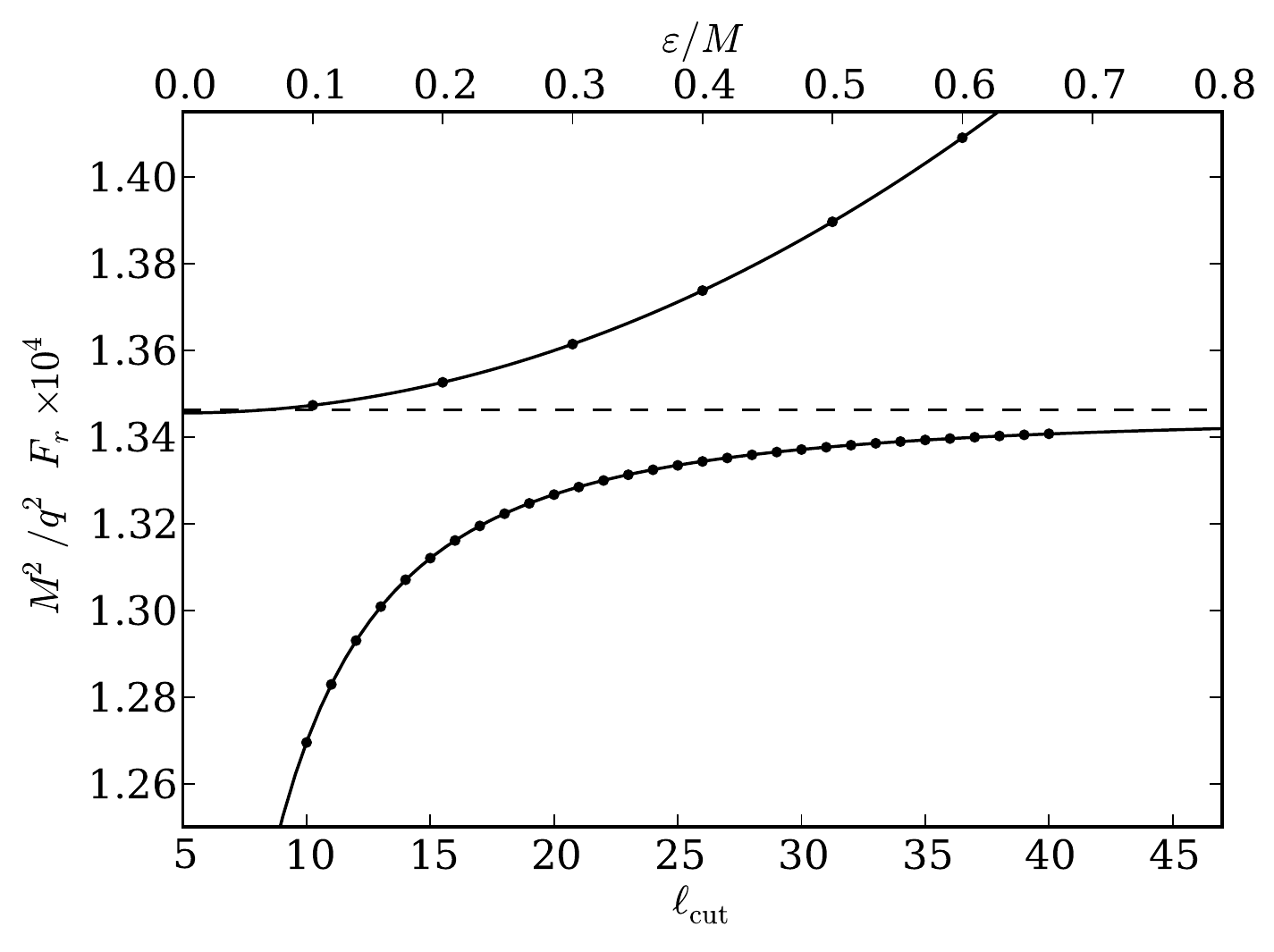}
 \caption{
Extrapolation in $\ell_{\rm cut}$ (bottom curve) and $\varepsilon$ (top curve) when computing the
radial component of the self-force for the $e=0.5$, $p=7.2$ eccentric orbit. The convergence in
both cases is approximately quadratic, as indicated by the curves, which are a least squares fit of
a quadratic function to the data. The dashed line indicates the high-accuracy value from
\cite{Warburton:2011hp}. 
 }
 	\label{fig:ell-sigma}
\end{figure}

With our chosen numerical parameters, the dominant source of error in our calculation comes from two
related issues: the fact that we only sum over a finite number of $\ell$ modes and that we use a
Gaussian of finite width $\varepsilon$. Both approximations limit the minimum length scale (both in
space and time) of the features that we can resolve. As shown in Fig.~\ref{fig:ell-sigma} the error
from these approximations converges away quadratically in $\varepsilon$ and $1/\ell_{\rm cut}$. By
repeating the calculation for a series of values of $\varepsilon$ and $\ell_{\rm max}$ one can
extrapolate to $\varepsilon=0$ and $\ell_{\rm cut} = \infty$ using standard Richardson
extrapolation.

The fact that the extrapolation in both parameters can be done reliably allows for significant
improvements in the accuracy of the numerical results. However, doing so requires multiple
numerical simulations, one for each value of $\varepsilon$. Since the focus of this paper is
not on achieving maximal accuracy but rather on demonstrating the feasibility of the method,
we do not extrapolate in $\varepsilon$ for the results in Sec.~\ref{sec:results} and instead
rely on extrapolation in $\ell$ alone. Although this does not yield the same accuracy as
additionally extrapolating in $\varepsilon$ (which gives an additional factor of 10 improvement
in accuracy in our test cases), it does nevertheless improve the accuracy relative to the
un-extrapolated result by a factor of 10.

\subsubsection{Numerical evolution}
For the Schwarzschild spacetime, it is convenient to choose coordinates that locally match the
Regge-Wheeler form in the vicinity of the worldline, but which also map future null infinity and the future
event horizon to finite values. This keeps the construction of initial data and the interpretation
of results simple, while improving computational efficiency by eliminating problems with outer boundaries. 
To this end, we employ hyperboloidal compactification \cite{Zenginoglu:2007jw} in the form of a layer
\cite{Zenginoglu:2010cq} with the standard Schwarzschild time and tortoise coordinates in the vicinity of 
the worldline and hyperboloidal coordinates far from the worldline using the double layer construction
from \cite{Bernuzzi:2012ku}.

We choose a compactifying radial coordinate $\rho$ defined through $r_\ast = \rho/\Theta(\rho)$, where
$\Theta$ is a function that is unity in a compact domain and smoothly approaches zero on both ends of 
the domain. The hyperboloidal time transformation is determined such that
$t + r_* = \tau + \rho$ for $r_\ast \to -\infty$, and 
$t - r_* = \tau - \rho$ for $r_\ast \to \infty$, which implies that level sets of the new time coordinate $\tau$ 
are horizon-penetrating near the black hole and hyperboloidal near null infinity. See \cite{Bernuzzi:2012ku}
for details of this double hyperboloidal layer construction.

Substituting the spherical-harmonic decomposition of the field, Eq.~\eqref{eq:field-Legendre}, into
the homogeneous wave equation, Eq.~\eqref{eq:homogeneous-wave}, we obtain $1$+$1$D equations for each
of the $\Psi_{\ell}$ which can be written in first order in time form,
\begin{align}
\label{eq:mol1}
\partial_\tau \Psi_\ell =& \frac{\Pi_\ell}{1\pm H} - \frac{H \partial_\rho \Psi_\ell}{1\pm H}\,, \\
\partial_\tau \Pi_\ell =& \partial_\rho \Big[\frac{\partial_\rho \Psi_\ell}{1\pm H} - \frac{H\, \Pi_l }{1\pm H} \Big] \nonumber \\
 & - (\Theta - \rho\, \Theta') \Big[1-\frac{2M \Theta}{\rho}\Big] \Big[\frac{\ell(\ell+1)}{\rho^2}+\frac{2M\Theta}{\rho^3} \Big]\Psi_\ell\,,
\label{eq:mol2}
\end{align}
where a prime denotes differentiation with respect to $\rho$,
$\Pi_\ell \equiv (1\pm H)\partial_\tau \Psi_\ell + H \partial_\rho \Psi_\ell$, and the function $H$ is given by
$H \equiv \pm \tfrac{d}{d r_\ast}(r_\ast - \rho) = \pm \left(1-\tfrac{\Theta^2}{\Theta - \rho \Theta'}\right)$, where the undetermined signs are positive near null infinity and negative near the black hole 
\cite{Bernuzzi:2012ku}. 

We evolve \eqref{eq:mol1} and \eqref{eq:mol2} for $200$ $\ell$ modes in the time domain using the
method-of-lines formulation. We compute spatial derivatives 
using eighth-order accurate centered finite differencing. At the boundaries (corresponding to the
black-hole horizon and null infinity) we use eighth-order asymmetric stencils without imposing any
boundary conditions because all characteristics are purely outgoing in our hyperboloidal coordinates.
We include a small amount of Kreiss-Oliger \cite{Kreiss-Oliger} numerical 
dissipation to damp out high-frequency noise. We evolve forwards in time using
standard fourth order Runge-Kutta time integration with a fixed step size equal to half of the
spatial grid spacing. We use an equidistant spatial grid with a resolution of $\Delta r^\ast = 0.01M$ 
in the  Regge-Wheeler-Zerilli coordinate region. We choose a domain of 
$\rho \in [-150M,150M]$ with layer interfaces located at $\pm 135M$, so that 
all worldlines we consider are moving in the Regge-Wheeler-Zerilli coordinate region.

We set Gaussian initial data of width $\varepsilon = 0.1M$. We evolve  for a
total time of $t=400M$, so that the remaining portion of the history integral
has a negligible contribution to the self-force and its contribution to the self-field
is well approximated by the late-time asymptotics described in Sec.~\ref{sec:tail}.
We extract values on the worldlines by simultaneously evolving the
equations of motion using the osculating orbits framework \cite{Pound:2007th} for the circular and
eccentric orbits and using \eqref{eq:radial-geodesic} for the radial infall case. 
We interpolate the values on the worldline using eighth order Legendre polynomial interpolation.

\subsection{Analytical approximations to the retarded Green function}
\label{sec:analytical}

In this section we describe the two analytical approximations to the RGF, one valid at
early times and one valid at late times, which we use as a substitute for the numerical
solution in the regions where it cannot be used. 

\subsubsection{Quasi-local expansion}
\label{sec:quasilocal}

In the quasi-local region, the spacetime points $x$ and $x'$ are assumed to be
sufficiently close together that the RGF is uniquely given
by the Hadamard parametrix, Eq.~\eqref{eq:Hadamard}. The term involving
$U(x,x')$ does not contribute to the integrals in Eqs.~\eqref{eq:Phisym1} and \eqref{eq:Fsym1}
since, within those integrals, it
has support only when $x=x'$, and the integrals exclude this point. We will
therefore only concern ourselves in the quasi-local region with the calculation
of the function $V(x,x')$.

The proximity of $x$ and $x'$ implies that an expansion of
$V(x,x')$ in the separation of the points can give a good
approximation within the quasi-local region. Reference~\cite{Casals:2009xa} used a WKB
method to derive such a coordinate expansion and gave estimates of its range of
validity. Referring to the results therein, we have $V(x,x')$ as a power series
in $(t-t')$, $(1-\cos \gamma)$, and $(r-r')$,
\begin{equation} \label{eq:CoordGreen}
V(x,x') =
 \sum_{i,j,k=0}^{\infty} v_{ijk}(r) ~ (t-t')^{2i} (1-\cos\gamma)^j (r-r')^k,
\end{equation}
where $\gamma$ is the angular separation of the points 
and the $v_{ijk}$ are computable analytic functions of $r$. 
Up to an overall minus sign, Eq.~\eqref{eq:CoordGreen} gives the quasi-local contribution to
the RGF. It is straightforward to take partial derivatives of these expressions at either
spacetime point to obtain the derivative of the Green function.

As proposed in Ref.~\cite{Casals:2009xa}, we use a Pad\'e
re-summation of Eq.~\eqref{eq:CoordGreen} in order to increase the accuracy and extend the domain of validity
of the quasi-local expansion. While not essential, this increases the region of overlap between the quasi-local and 
numerical Gaussian domains. Since Pad\'e re-summation is only well defined for series expansions
in a single variable, it is necessary to first expand $r'$ and $\gamma$ in a Taylor series in
$t-t'$, using the equations of motion to determine the higher derivatives appearing in the series
coefficients. Then, with $V(x,x')$ written as a power series in $t-t'$ alone, a standard
diagonal Pad\'e approximant of order $(t-t')^{26}$ in both the numerator and denominator is
constructed and used to represent the Green function in the quasi-local region.

\subsubsection{Late-time behavior}
\label{sec:tail}

It is known since the work of Price~\cite{Price:1971fb,Price:1972pw} that an initial field perturbation of a Schwarzschild black hole decays at late times
with a leading power-law behavior: $t^{-2\ell-3}$ for the field mode with multipole $\ell$.
This late-time behaviour is related to the form  at large radius of the  potential in the ordinary differential equation satisfied by the radial part of the perturbation, i.e.,
the Regge-Wheeler equation for general integer spin of the field.
The form of the late-time tail can be derived from low-frequency asymptotics along the branch cut that the Fourier modes of the Green function 
possess in the complex-frequency plane~\cite{Leaver:1986gd,Ching:1995tj}.
In~\cite{Casals:2012tb}, the exact coefficients in the expansion at late times of the Green function modes at arbitrary radius were given up to the first four orders.
In particular, a new behavior $t^{-2\ell-5}\log (t/M)$ was obtained in the third-order term, thus showing a deviation from a pure power-law expansion.
In the present paper we have used the method presented in~\cite{Casals:2012tb}, which is described in detail in~\cite{Casals:Ottewill:2014}.
This method is based on the `MST formalism'~\cite{mano1996analytic,Mano:1996vt,Sasaki:2003xr}.
It consists of expressing the  solutions to the radial ordinary differential equation as series of special functions,
namely ordinary hypergeometric functions and confluent hypergeometric functions. These series have the advantage that they are  particularly amenable to
expansions for low frequency. We have used these series in order to obtain low-frequency asymptotic expansions
up to next-to-leading order of the Green function modes along the branch cut.
Finally, by integrating these expansions along the branch cut we obtain the late-time asymptotics of the full Green function, which we will 
to as the late-time tail.

\section{Results}
\label{sec:results}

\begin{figure}
 \includegraphics[width=0.85\columnwidth]{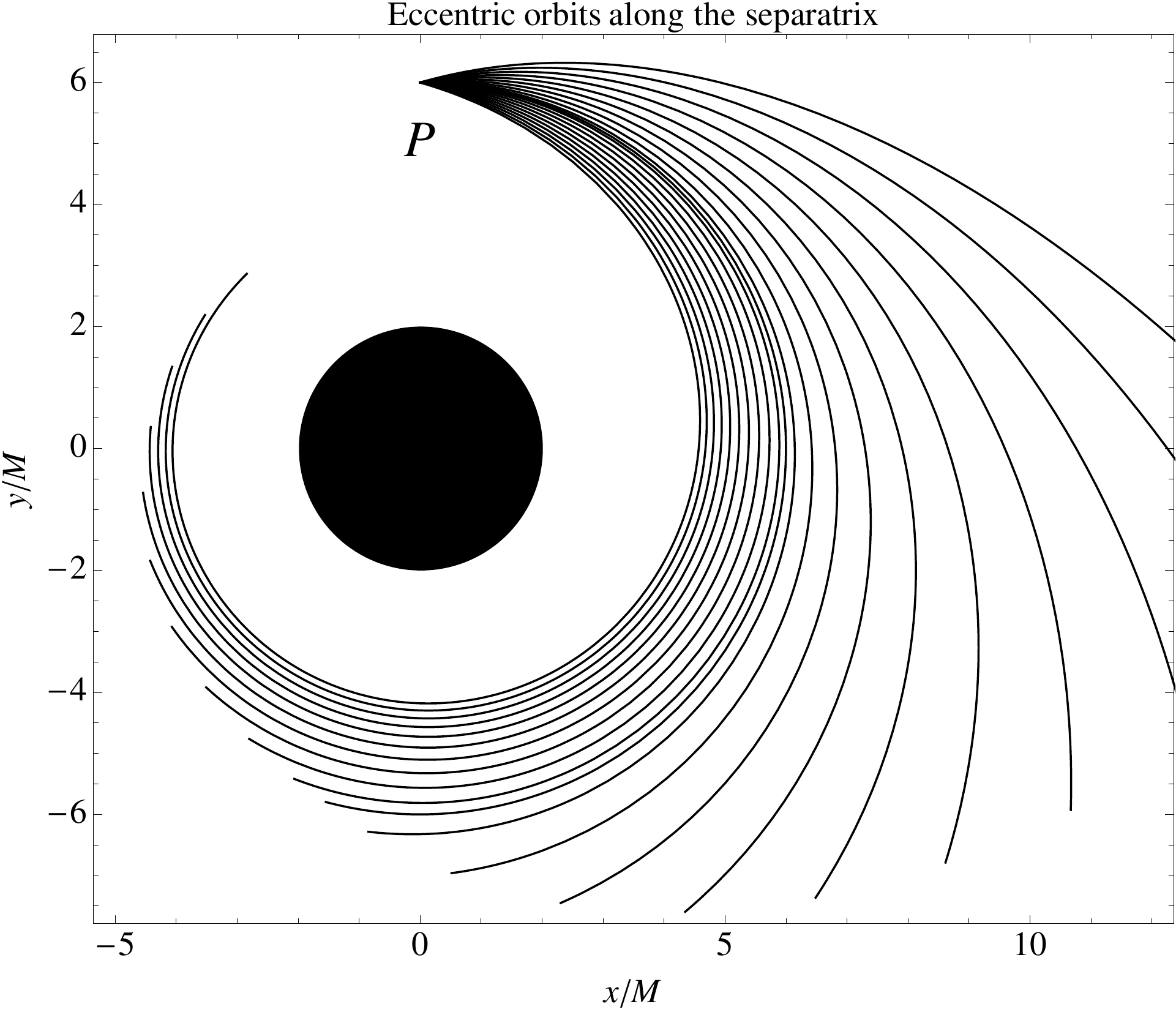}
 \caption{
A sample of bound geodesic orbits having different eccentricities but passing through the same
point $P$ at $r_0 = 6M$. For any of these geodesics, the history-dependent part of the self-field and self-force at $P$
can be computed from four numerical time-domain evolutions (i.e., $G$, $\partial_t G$, $\partial_r G$, and $\partial_\varphi G$) with $P$ as the base
point. The same is true for {\it any} worldline passing through $P$, even accelerated or
unbound ones. 
 }
 	\label{fig:geodesics}
\end{figure}

In this section we present self-force computations for a variety of orbits at
the point where the orbits pass through $r_0=6M$. One of the strengths of the Green function approach 
is that once the Green function and its derivatives are available, the self-force for all orbits passing through 
one point can be computed by simple worldline evaluations (see Fig.~\ref{fig:geodesics}). The entire set of
results in this section are obtained from just three numerical time-domain evolutions: for the Green
function and its time and radial derivatives (because of the spherical-harmonic decomposition,
angular derivatives act only on the Legendre polynomials and the numerical data required is the same
as that for the undifferentiated Green function).

Here, we focus on three particular families which best demonstrate the flexibility of the method:
\begin{itemize}
\item Accelerated circular orbits, including ultra-relativistic and static limits.
\item Geodesic eccentric orbits, including those near the separatrix between stable and unstable bound geodesics.
\item Radial infall, reflecting the ability to handle unbound motion.
\end{itemize}
Before discussing these cases, we first validate our methods and establish the 
accuracy of our results by comparing against two cases for which accurate self-force values are available.

\subsection{Validation against existing results}
\label{sec:validation}

The frequency-domain based mode-sum regularization method is unparalleled in its ability to compute
highly accurate values for the self-force in cases where the retarded field is given by a discrete
spectrum involving a small number of Fourier modes
\cite{Haas:2006ne,Detweiler:2002gi,Warburton:2011hp,Heffernan:Ottewill:Wardell}. Here,
we consider two cases which were computed to high accuracy using the methods in
\cite{Warburton:2011hp,Heffernan:Ottewill:Wardell} and which were considered in~\cite{Casals:2013mpa}: a circular geodesic orbit of radius $r_0 = 6M$,
and an eccentric orbit with eccentricity $e=0.5$ and semilatus rectum $p=7.2$.

Our numerical calculations employ a number of approximations:
\begin{itemize}
  \item Numerical discretization of the $1$+$1$ dimension equation. This introduces a numerical error which
        arises from errors in the time integration, spatial finite differencing, and interpolation of
        values to the worldline, all of which converge away with increasing resolution. Using
        our numerical grid with spacing $\Delta r^\ast = 0.01$M, these errors are all negligible.
        For example, in the most difficult case of the radial component of the self-force for the
        eccentric orbit, a higher resolution simulation with $\Delta r^\ast = 0.005$M was used to
        estimate that the relative error from numerical discretization is $\sim 1 \times 10^{-5}$.
  \item Taylor series approximation at early times. The accuracy of the Taylor series
        approximation decreases as the field and source points are separated. By repeating the
        calculation for a range of matching times (between the Taylor series and the numerical solution) 
        in the region $t_m\in[15M,21M]$, we
        obtained relative differences in the radial self-force for the eccentric case of
        $\sim 10^{-3}$. Although these differences are comparable to those arising from a finite
        cut-off in the sum over $\ell$, it turns out that this is because the two issues are closely
        linked: by using a larger cutoff in the $\ell$-sum, the errors from matching are also
        reduced. This connection also allows us to use the error from the finite cut-off in the
        $\ell$-sum as an accurate approximation for the error from matching by choosing the matching
        time so that the error is no larger than that from the $\ell$-sum.
  \item Asymptotic expansions for the late-time tail. These expansions become increasingly unreliable as
        field and source points are brought closer together. By numerically evolving to $t=400M$,
        we make sure that the remainder of the history-integral contributes a negligible amount to the self-force. 
        In our test eccentric orbit case, the relative contribution is $\sim 5 \times 10^{-5}$. \\ The
        contribution to the history integral from late times is more significant for the scalar
        field because the Green function decays more slowly in time than its derivative.
        However, even in this case the magnitude of the late-time tail history integral for the
        circular orbit is only
        $\sim 10^{-5}$ (compared to $\sim 10^{-2}$ for the entire history integral) and
        any error in our approximation is considerably smaller. We may therefore neglect
        errors from the late-time portion of the history integral. Unfortunately, this
        does not hold for generic orbits. For example, for the eccentric orbit test
        case, the late-time-tail-integral error slightly dominates over other sources of error in
        the calculation; this dominance becomes more and more pronounced for orbits whose radial
        position varies more and more unpredictably with time. For this reason, when quoting errors
        for the self-field, we use the magnitude of the late-time tail integral as an estimate
        under the assumption that it always provides an upper-bound (if somewhat
        over-conservative) approximation.
  \item Replacement of a delta distribution with a Gaussian. This causes both a spurious
        pulse at early times and a smoothing out of any sharp features. The initial pulse is
        eliminated by replacing the Green function at early times with its quasilocal Taylor
        series expansion. With our chosen Gaussian width of $\varepsilon = 0.1M$, the relative
        error caused by the finite-width Gaussian is $\sim 1 \times 10^{-3}$ (see Fig.~\ref{fig:ell-sigma}).        
  \item Cutting off the infinite sum over $\ell$. Like the finite width Gaussian, this has the
        effect of smoothing out of any sharp features. We  estimate this error by the difference
        between the self-force computed using $\ell_{\rm cut} = 40$ and the value obtained by
        extrapolating the curve in Fig.~\ref{fig:ell-sigma}, which gives a relative
        error $\sim 4 \times 10^{-3}$.
\end{itemize}
In our test cases, the dominant source of error therefore comes from the choice of
$\varepsilon$ and the cutoff in the sum over $\ell$. Because the two sources of error are intimately
connected (a choice of $\varepsilon$ results in an effective maximum $\ell$ which can
be resolved) we estimate the error in our results by considering the errors in
the sum over $\ell$ (for sufficiently small $\varepsilon$). This estimate is conservative 
because it ignores accuracy improvements from extrapolating in $\ell_{\rm cut}$. 
The true accuracy of the computed self-force is up to an order of magnitude better 
(see Table \ref{tab:errors}).

\begin{table}
\begin{tabular}{r||c|r|c|c}
\hline \hline
 & & Computed value & Rel. Err. & Est. Err. \\
\hline
\multirow{4}{*}{\rotatebox{90}{Circular}} & $M/q \, \Phi$ & $-5.45517 \times 10^{-3}$ & $6 \times 10^{-5}$ & $3 \times 10^{-3}$ \\
& $M^2/q^2 \, F_t$ & $3.60779 \times 10^{-4}$ & $4 \times 10^{-4}$ & $2 \times 10^{-3}$ \\
& $M^2/q^2 \, F_r$ & $1.67861 \times 10^{-4}$ & $8 \times 10^{-4}$ & $2 \times 10^{-3}$ \\
& $M^2/q^2 \, F_\varphi$ & $-5.30452 \times 10^{-3} $ & $5 \times 10^{-5}$ & $5 \times 10^{-4}$ \\
\hline
\multirow{4}{*}{\rotatebox{90}{Eccentric}} & $M/q \, \Phi$ & $-7.70939 \times 10^{-3}$ & $1 \times 10^{-3}$ & $1 \times 10^{-3}$ \\
& $M^2/q^2 \, F_t$ & $6.65241 \times 10^{-4}$ & $2 \times 10^{-4}$ & $1 \times 10^{-3}$ \\
& $M^2/q^2 \, F_r$ & $1.3473 \times 10^{-4}$ & $8 \times 10^{-4}$ & $4 \times 10^{-3}$ \\
& $M^2/q^2 \, F_\varphi$ & $-7.28088 \times 10^{-3}$ & $4 \times 10^{-5}$ & $5 \times 10^{-4}$ \\
\hline
\end{tabular}
\caption{Numerical results for circular and eccentric orbit test cases, including estimated errors.}
\label{tab:errors}
\end{table}

In Table~\ref{tab:errors} we compare the results of our numerical Green-function calculation with
reference values computed using the frequency-domain mode-sum regularization method. We also give
internal error estimates based on the assumption that the choice of a finite $\ell_{\rm cut}$
reasonably reflects the dominant source of error in the self-force and the finite integration time
is the dominant source of error in the self-field. Our goal here is not to show that the Green
function method improves on the accuracy of existing methods. Frequency-domain based methods
are the accuracy leaders for cases where they can be used. But the Green function method provides a
highly flexible and complementary approach which gives good results using modest computational
resources. Furthermore, it would not be difficult to significantly improve on the accuracy of the results
presented here through either brute force methods (higher resolution, more $\ell$ modes, smaller
Gaussian, higher order Taylor series and longer integration times) or through readily available 
improvements to the numerics (better hyperboloidal coordinates, spectral methods for spatial derivatives, 
improved time-integration  schemes and analytic asymptotics for the large-$\ell$ 
modes~\cite{Nolan-Capra2013,Casals:2012px}). We leave the implementation of such improvements 
for future work.

\subsection{Accelerated circular orbits}
\label{sec:accelerated}
We consider a particle in a circular orbit of radius $r_0$ and constant angular velocity $\Omega$, so that
the azimuthal angle coordinate is given by $\varphi=\Omega \, t$. For such orbits, the redshift factor is 
\begin{align}
	z \equiv \frac{1}{u^t} = \sqrt{1-\frac{2M}{r_0} - r_0^2 \Omega^2} .
\label{eq:redshift1}
\end{align}
The three special cases $\Omega^2=\{0,M/r_0^3,(r_0-2M)/r_0^3\}$ correspond
to a static particle, circular geodesic, and null orbit, respectively. Because the orbit is
accelerated, there is an additional instantaneous contribution to the self-force not present for a
geodesic. This instantaneous contribution is given by \cite{Quinn:1996am} 
\begin{align}
	F^{\rm inst}_\mu = \frac{q^2}{3} (g_\mu{}^\nu + u_\mu u^\nu) \frac{ Da_\nu}{ d\tau}  \, ,
\end{align}
which, in the constantly accelerated circular orbit case, has components
\begin{gather}
  F^{\rm inst}_t = \frac{q^2 M^2 (r_0 \!-\! 2M) (n^2 \!-\! 1)}{3 r_0^5 z^3} \bigg[ 1 \!+\! \frac{ (r_0 \!-\!2M) (n^2\!-\!1)}{ r_0 z^2} \bigg],
\nonumber \\
  F^{\rm inst}_r = 0, \quad F^{\rm inst}_\theta = 0,
\nonumber \\
  F^{\rm inst}_\varphi = - \frac{q^2M^{3/2} (r_0 \!-\! 2M) n (n^2 \!-\! 1)}{3r_0^{7/2} z^3} \bigg[ 1 \!+\! \frac{M(n^2\!-\!1)}{ r_0 z^2}  \bigg],
\end{gather}
where $n \equiv \Omega / \Omega_g$, and $\Omega_g = (M/r_0^3)^{-1/2}$ is the geodesic frequency.

\begin{figure*}
	\includegraphics[width=2\columnwidth]{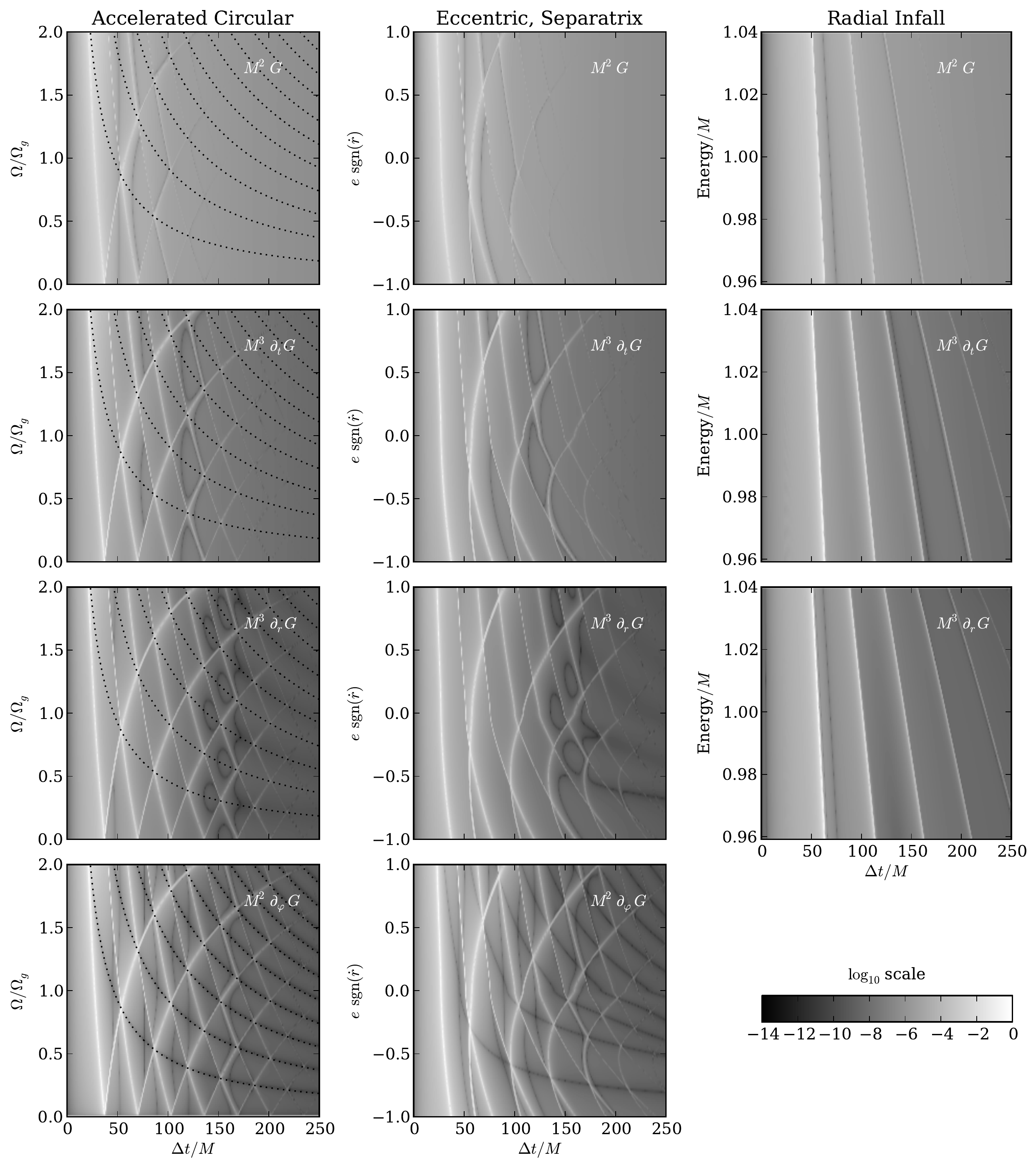}
	\caption{Plot of the Green function and its derivatives for all orbits considered in this paper,
evaluated at the base point where the orbital radius is $r_0=6M$. The $x$-axis labels time $\Delta
t/M$ from coincidence and the $y$-axis labels the parameter defining the orbit. The dotted curves
for accelerated circular orbits correspond to the times at which the orbit crosses a point
with an angle $\varphi = n\pi$, $n\in\mathbb{Z}^+$ relative to the Green function's base point at $t=0$.
Note: since light-shaded curves effectively correspond to null geodesics, the particular space-time points where two light-shaded curves intersect 
must necessarily lie, by symmetry, on a dotted curve (i.e., with $\varphi = n\pi$) and they are the caustic points which are encountered
by the timelike orbits that we consider.
}
	\label{fig:G}
\end{figure*}

We compute the RGF and its derivatives on
all constantly accelerated circular orbits with radius $r_0 = 6M$, and plot the results in
Fig.~\ref{fig:G} as a function of the time  $\Delta t / M$ from coincidence and of the orbital
frequency $\Omega$ relative to the geodesic value. The figure shows the complex structure of the
Green function that results from the interplay of wave propagation with the orbiting particle; the
light shaded regions correspond to caustic echoes along the orbits. The dotted curves indicate
the times at which a given circular orbit crosses a
point with  $\varphi = n\pi$ for $n$ a positive integer.

The left column in Fig.~\ref{fig:G} shows the Green function itself (top left) and its derivatives (bottom three) evaluated along the
orbits. As the shading in the plots indicate, the 
late-time tail decay is slowest for the Green function, faster for its time derivative, and fastest for the radial and azimuthal derivatives.
We thus infer that late-time effects from
back-scattering are more relevant for computing the self-field than they are for the self-force.
This is in agreement with what was observed in~\cite{Casals:2013mpa} in the specific instances of a circular geodesic and an eccentric orbit with $p=7.2$ and $e=0.5$.
Conversely, caustic echoes give non-negligible contributions to the self-force at later times than they would for the self-field because the late-time tail contributes less.
The contributions from the effects of curved geometry
(back-scattering) and trapping (caustic echoes) to quantities of interest are clearer in our
method than in other self-force calculation techniques.

\begin{figure}[h]
 \includegraphics[width=\columnwidth]{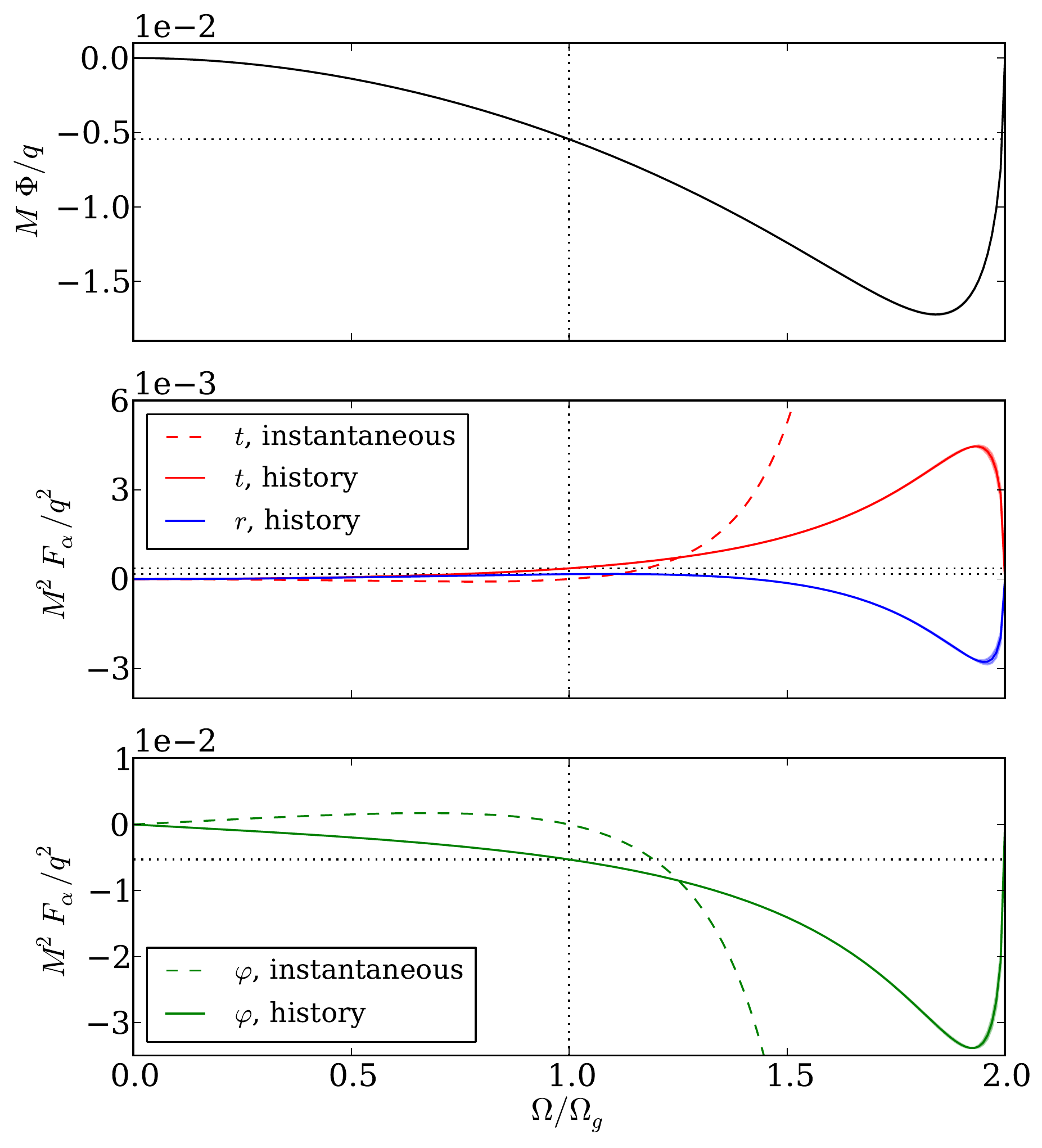}
 \caption{
Regular part of the self-field and self-force on constantly accelerated circular orbits of radius
$r_0=6M$ with different orbital frequencies $\Omega$ relative to the geodesic frequency $\Omega_g$.
The motion is null when $\Omega = 2 \Omega_g$. The dashed black lines indicate the reference value of
the self-field at the geodesic frequency from \cite{Haas:2006ne}. Error bars are included as shaded regions about the curves.
}
 	\label{fig:Phi-circ}
\end{figure}

The history-dependent part of the self-field and self-force evaluated along these orbits are given by
\begin{align} \label{eq:hist-circ}
  \Phi_{\rm hist} =& q \sqrt{1-\frac{2M}{r_0} - r_0^2 \Omega^2} \, \int^{0^-}_{-\infty} \!\!\! G_{\rm ret}(x; x') dt',
\nonumber \\
  F^{\rm hist}_\mu =& q^2 \sqrt{1-\frac{2M}{r_0} - r_0^2 \Omega^2} \,  \int^{0^-}_{-\infty} \!\! \partial_\mu G_{\rm ret}(x; x')  dt' .
\end{align}
These are plotted in the top panel of Fig.~\ref{fig:Phi-circ} as a function of $\Omega/\Omega_g$.

When the particle is at rest ($\Omega = 0$) the self-field and self-force are 
consistent with zero within our error bars, as expected and in agreement with analytical results
 \cite{Wiseman:2000rm}. Furthermore, the local and
history-dependent pieces of the self-force components vanish separately, independently of each
other. We also find that the history-dependent self-field and self-force vanish for a null
circular orbit, [$\Omega = (r_0-2M)/r_0^3$]; this can be inferred immediately from
Eq.~\eqref{eq:hist-circ} because the redshift factor,
\begin{align}
	z \big|_{\rm null~orbits} = \sqrt{ 1 - \frac{ 2 M}{r_0 } - \frac{ r_0 - 2M }{ r_0} }  = 0
\end{align}
vanishes for all orbital radii. This is expected in the ultra-relativistic regime because the charge-field 
interaction term in the action is $q \int dt \, \Phi (z^\alpha(t)) / \gamma$ where $\gamma$ 
is the boost factor relative to the given frame (e.g., the inertial frame of a distant observer). Because the 
field from a moving charge scales as $\Phi \sim q/(\gamma M)$, the scalar field amplitude decreases as 
the boost increases until $\Phi$ vanishes in the ultra-relativistic limit $\gamma \to \infty$. This situation is
different in gravity where the metric perturbations couple strongly to the motion of the small mass, 
$h_{\mu\nu} \sim \gamma m /M$, and care must be taken in defining the perturbation theory in the 
ultra-relativistic regime where $\gamma \gg 1$ \cite{Galley:2013eba}.

The history-dependent piece of the self-field on the circular orbits attains a minimum value of
approximately $-0.01722 \, q/M$ at $\Omega \approx 1.84 \, \Omega_g$ and a maximum value of $0$ in
the static and null limits. The maximum and minimum values of the history-dependent pieces of
$\{F_t, F_r, F_\varphi\}$ are approximately $\{ 0.004459, 0.0001731, 0\} \, q^2/M^2$ and
$\{0, -0.002786, -0.03387\} \, q^2/M^2$, respectively, at
frequencies $\Omega/\Omega_g \approx \{ 1.93, 1.08, 0 \text{ and } 2\}$ and $\Omega/\Omega_g \approx \{0 \text{ and } 2, 1.95, 1.93 \}$.

\begin{figure}[h]
 \includegraphics[width=\columnwidth]{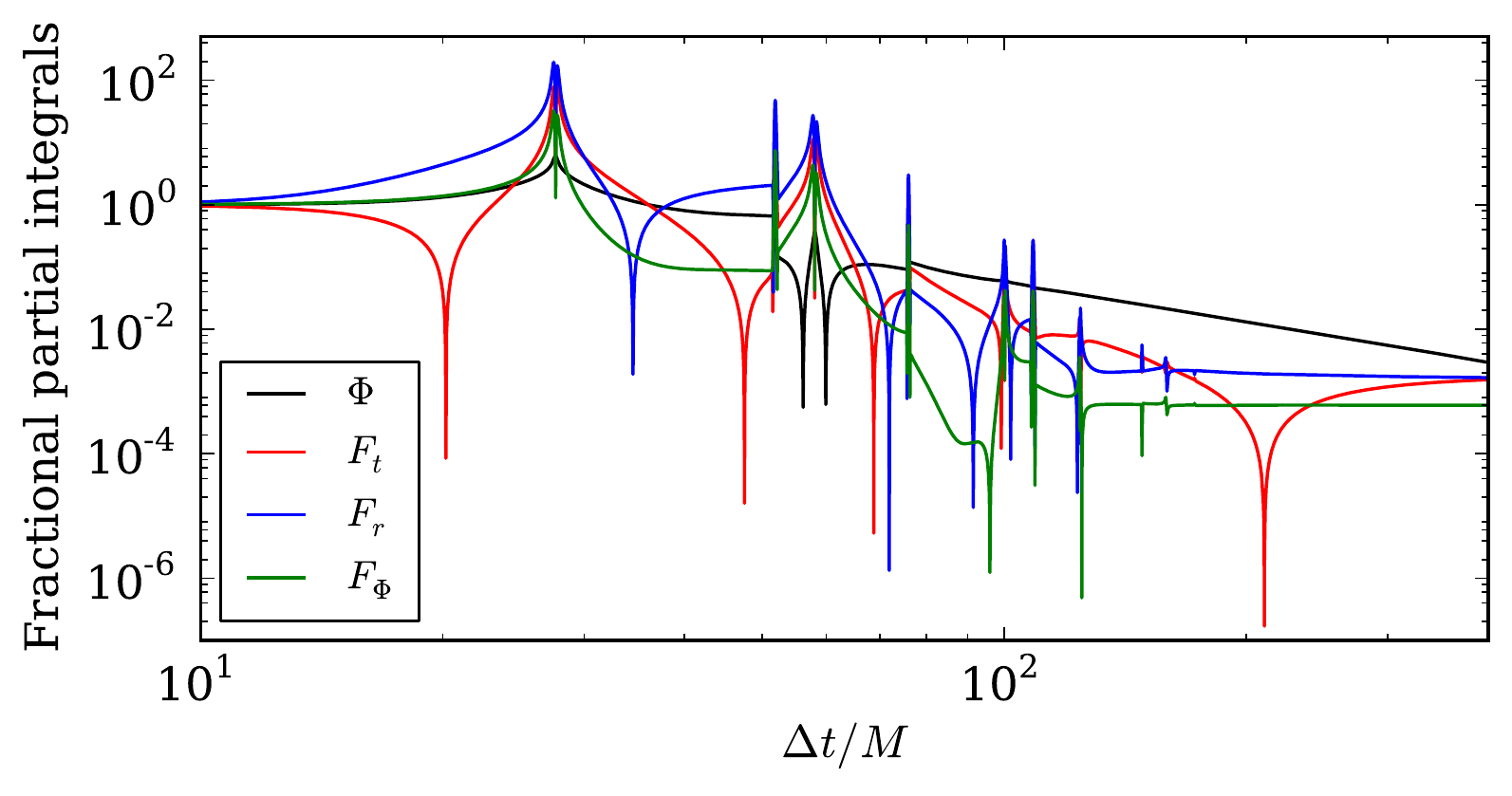}
 \caption{Fractional difference of the partial self-field and self-force components relative to the reference values in \cite{Haas:2006ne} for a circular geodesic orbit at $r_0=6M$. 
 See Fig.12 in~\cite{Casals:2013mpa} for the corresponding plots of the actual partial self-field and self-force.
 }
 	\label{fig:partial}
\end{figure}

The RGF allows us distinguish which aspects of wave propagation in black hole spacetimes (caustic
echoes or late-time backscattering) affect the self-field and self-force. Figure \ref{fig:partial} shows the partial 
self-field and the self-force components for the circular geodesic orbit at $r_0 = 6M$ relative to the reference 
values $ \Phi_{\rm reference}$ computed in \cite{Haas:2006ne}. The partial self-field is defined as the integral
in Eq.~(\ref{eq:hist-circ}) but with the lower limit of integration replaced by $-\Delta t$ as the independent
variable in the figure. The fractional partial self-field is defined as
\begin{align}
	\bigg| 1 - \frac{ \Phi_{\rm partial}(\Delta t) }{ \Phi_{\rm reference} } \bigg|.
\end{align}
Similar definitions are taken for the fractional partial self-force components.

Most of the sharp features in Fig.~\ref{fig:partial} are caustic echoes intersecting the circular orbit. After 
$\sim 80M$ caustic echoes no longer appreciably affect the self-field while this occurs for the self-force 
components only after $\sim150M$. This is roughly compatible with the structure of the RGF and its 
derivatives observed in the left column of Fig.~\ref{fig:G}. For this particular orbit we see from Fig.~\ref{fig:G} 
that there are $8$ or $9$ caustic echoes within $\sim 150M$. The late-time tail from 
backscattering appears to have little influence, if any, on the radial and azimuthal self-force components 
as indicated by the plateaus after $\sim 150M$. However, the self-field and the time component of the self-
force seem to depend more strongly on the late-time tail. In fact, the difference between the partial integral 
for $F_t$ and the reference value changes sign just after $200M$, as can be seen in Fig.~\ref{fig:partial} 
by the downward pointing spike. As a result, the partial $F_t$ value does not converge to its final value 
until late in the integration. The late-time tail itself has a strong influence on the self-field as the former 
changes the latter's value by more than a factor of $30$ from $80M$ to $400M$. In addition, the self-field 
seems to have not yet converged after $400M$.

\subsection{Geodesic eccentric orbits}
\label{sec:eccentric}

Circular orbits are too special to be representative. For example, for circular orbits
$\partial_r G_{\rm ret} (x,x') = \partial_{r'} G_{\rm ret} (x,x')$
and one can avoid the careful treatment discussed in Sec.~\ref{sec:Kirchhoff} for
computing the derivative of the Green function. To illustrate the flexibility of the RGF method, 
we now study a family of eccentric orbits, which are representative of generic orbits in Schwarzschild
spacetime.

Bound geodesics of  Schwarzschild spacetime are conveniently parametrized by their eccentricity,
$e$, and semi-latus rectum, $p$. These can be defined in terms of the radial turning points of the
orbit, the periastron ($r_{\rm min}$) and apastron ($r_{\rm max}$),
\begin{equation}
  p = \frac{2r_{\rm min} r_{\rm max}}{M(r_{\rm min}+r_{\rm max})}, \quad e = \frac{r_{\rm max} - r_{\rm min}}{r_{\rm max} + r_{\rm min}}.
\end{equation}
The stable bound orbits are those for which $p \ge 6 + 2 e$. The separatrix, defined by $p = 6 + 2e$
separates the stable and unstable orbits and corresponds to the limiting case where the orbit comes
in from apastron and spends an infinite amount of time whirling around periastron. In terms of this
$p$--$e$ parametrization, the redshift factor for eccentric orbits is given by
\begin{equation}
  z = \left(1-\frac{2M}{r}\right)\sqrt{\frac{p (p - 3 - e^2)}{(p - 2 - 2 e) (p - 2 + 2 e)}}.
\end{equation}

We focus on the portion of the $p$--$e$ parameter space
corresponding to points along the separatrix. Fixing $p = 6+2e$ and $r_0 = 6M$ we parametrize the
orbits by their eccentricity $0 \le e \le 1$ along with the choice of whether the radial motion is
oriented inwards or outwards at $r_0 = 6M$. This family of geodesics are illustrated in
Fig.~\ref{fig:geodesics}.

\begin{figure}[h]
	\includegraphics[width=\columnwidth]{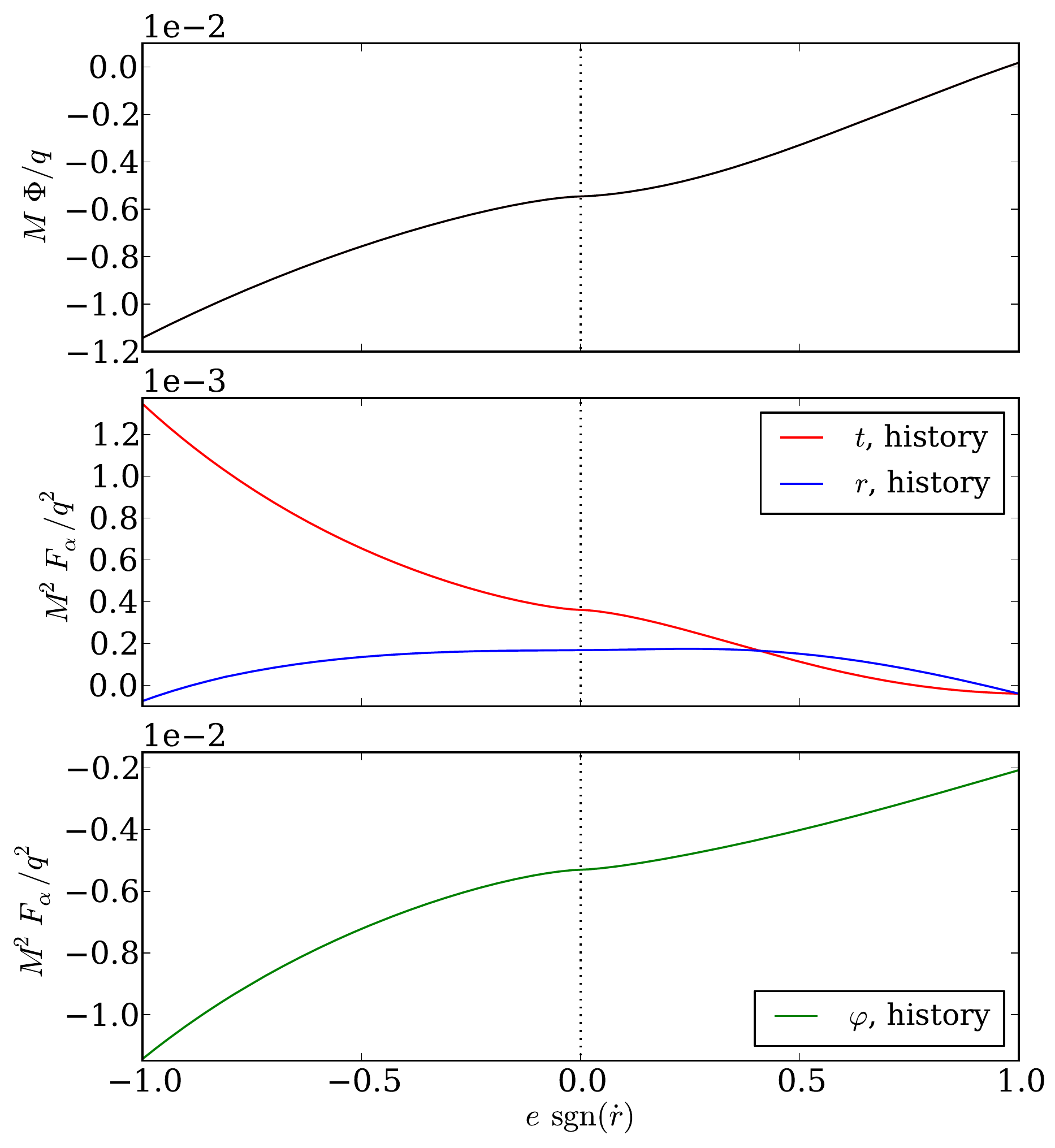}
 \caption{
Regular part of the self-field and self-force on eccentric geodesic orbits along the separatrix $p=6+2e$ at the
point when the orbital radius is $r_0=6M$. The orbits are parametrized by the eccentricity $e$
along with whether the motion is radially inward or outward, as measured by $\sgn \dot{r}$.
}
 	\label{fig:Force-ecc}
\end{figure}

In Fig.~\ref{fig:G} we plot the Green function (top, center) and its derivatives (bottom three,
center) along the orbit for this family of geodesics. We find similar features to the accelerated
circular orbits case, but the locations of the features are deformed as a result of the more complex
orbital shape. This gives rise to qualitatively different results for the self-field and self-force,
as illustrated in Fig.~\ref{fig:Force-ecc}. In this case $\{\Phi, F_t, F_r, F_\varphi\}$ attain
maximum and minimum values of approximately $\{ 0.0001839, 0.001347, 0.0001747, -0.002075\} \, q^2/M^2$
and $\{ -0.01143, -0.00004040, -0.00007567, -0.01144\} \, q^2/M^2$, respectively,
at eccentricities $e\sgn\dot{r} \approx \{ 1, -1, 0.24, 1\}$ and $e \sgn\dot{r}  \approx \{ -1, 1, -1, -1\}$.

\subsection{Radial infall}
\label{sec:infall}

As a final example, we consider the unbound motion of a worldline falling radially inwards and
compute the self-force at $r_0 = 6M$. Parametrizing the space of worldlines by the maximum radius  $r_{\rm max}$
they attain, the position $r(t)$ at some time $\Delta t$ in the past is given by
\begin{equation}
\label{eq:radial-geodesic}
   \ddot{r} - \frac{2M}{r(r-2M)} \dot{r}^2 + \frac{M(r-2M)^2}{r^4}\frac{r_{\rm max}}{(r_{\rm max}-2M)} = 0,
\end{equation}
where an overdot denotes differentiation with respect to Schwarzschild coordinate time. Once the
motion reaches its maximum radius, $r_{\rm max}$, we hold its position fixed at $r=r_{\rm max}$
for the remainder of the time. This corresponds physically to a particle being held at
rest at $r=r_{\max}$, then released and allowed to fall into the black hole.

\begin{figure}
 \includegraphics[width=\columnwidth]{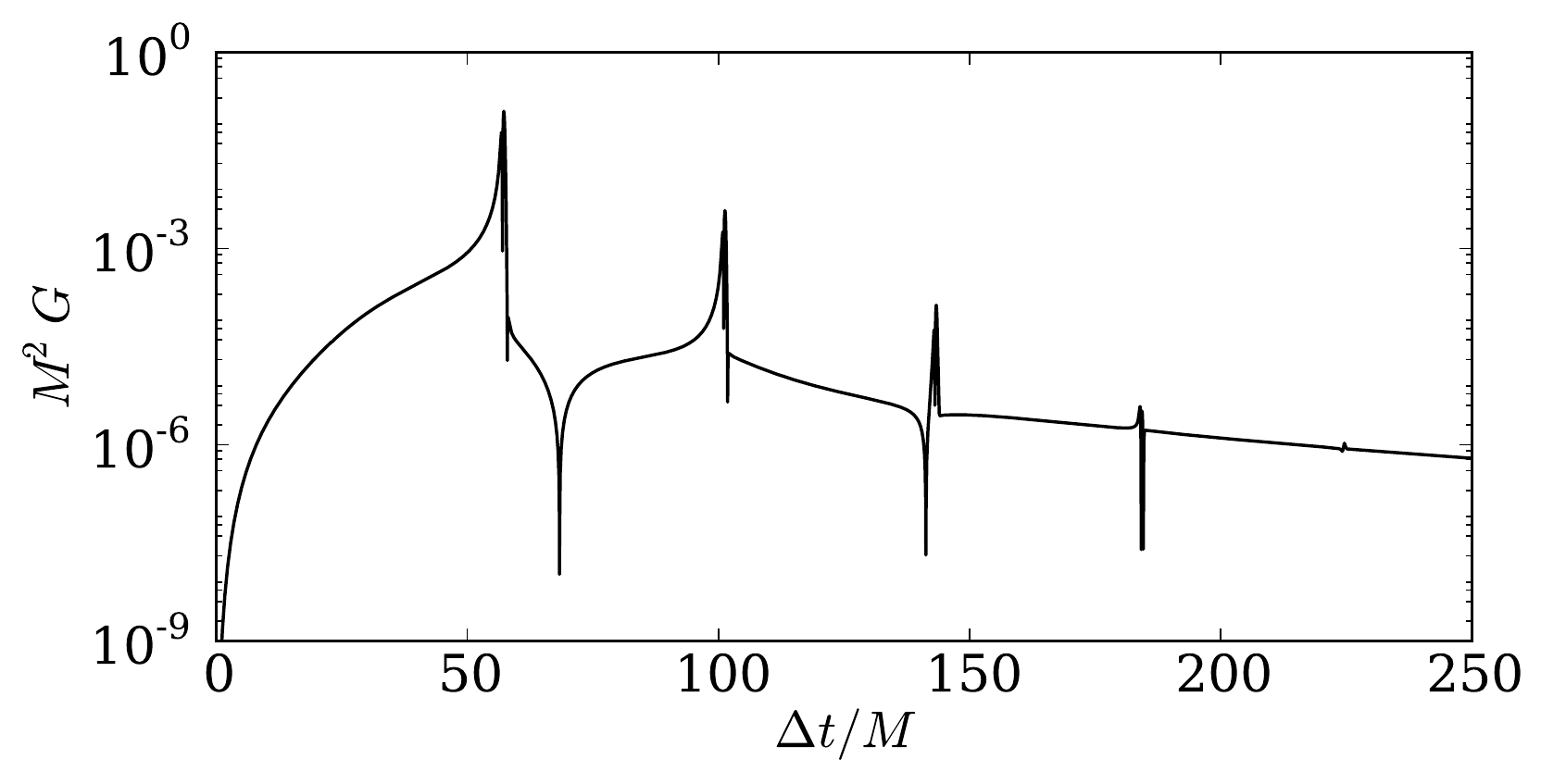}
 \caption{
Green function along the worldline corresponding to a radial geodesic starting from rest at $r_{\rm
max} = \infty$ and falling into $r_0=6M$. The two-fold singularity structure apparent in this case
is in contrast to the standard four-fold structure which is typically seen for orbits in the
Schwarzschild spacetime. 
 }
 	\label{fig:G-rad-static-infinity}
\end{figure}
In Fig.~\ref{fig:G} we plot the Green function along the worldline for this family of geodesics.
One particularly interesting feature of this case is that the Green function no longer has the
familiar four-fold structure of other orbits in Schwarzschild spacetime, but instead has a two-fold
structure. This is due to the fact that, in this case, the singularities in the Green function always happen at
caustics,
 where there is a $1$-parameter family of null geodesics crossing the worldline,
rather than just a single null geodesic. This two-fold structure is clearly illustrated in
Fig.~\ref{fig:G-rad-static-infinity}, showing the case $r_{\rm max} = \infty$.

\begin{figure}
	\includegraphics[width=\columnwidth]{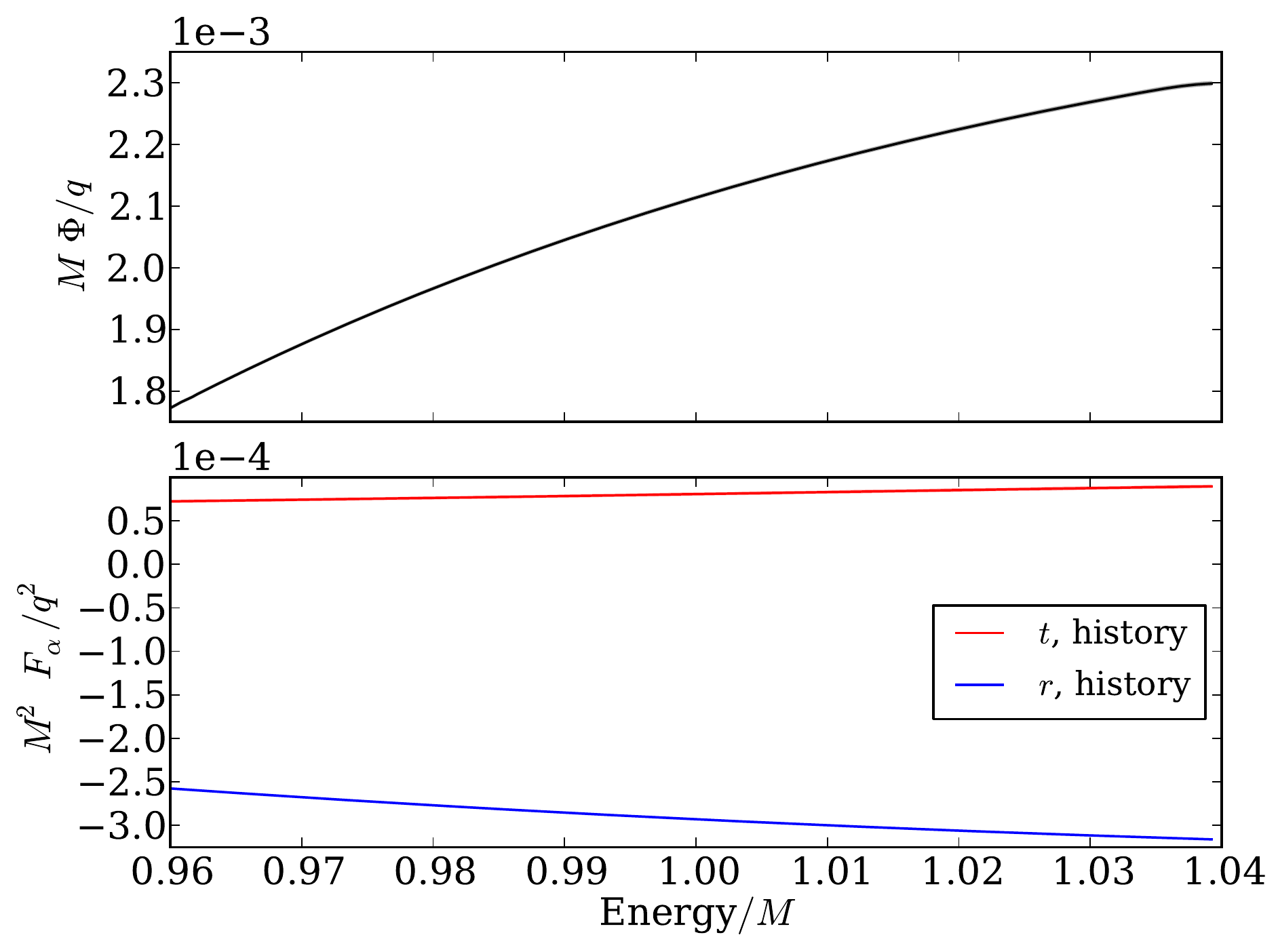}
 \caption{
Regular part of the self-field and self-force on radial infall worldlines released from rest at different
initial radii $r_{\rm max}$. The magnitude of the self-interaction increases as the energy of the
motion is increased by increasing $r_{\rm max}$. The extrapolation to $r_{\rm max} = 6M$ would
correspond to the static case, where we would expect the self-interaction to be exactly $0$.
 }
 	\label{fig:Phi-rad-static}
\end{figure}
In Fig.~\ref{fig:Phi-rad-static} we plot the self-field and self-force as a function of
the energy of the geodesic (which is directly related to $r_{\rm max}$). The symmetry of
the problem demands  $F_\varphi=0$ in all cases. Both the self-field and self-force
are monotonic functions of the energy, with larger (in magnitude) values for more energetic
worldlines. This behavior agrees with intuition; more energetic geodesics are moving faster and one could
reasonably expect stronger self-interaction as a result.

\section{Concluding remarks}
\label{sec:conclusion}

We have presented a new method for numerically computing the self-force based on a global
approximation of the RGF. The method takes advantage of the fact that the RGF is the fundamental
solution to the wave equation in a curved background and can be used to build any inhomogeneous
solution through a straightforward convolution integral with the source of interest. In this way,
the quintessential features of wave propagation are disentangled from the arbitrary source.

Global numerical approximations for the RGF can be obtained in at least two ways: either by
approximating the delta-distribution source by a Gaussian (as introduced in
\cite{Zenginoglu:2012xe}) or using a Kirchhoff representation and approximating delta-distribution
initial data by a Gaussian. In this paper we chose the latter and exploited the spherical symmetry
of the Schwarzschild background to numerically solve the initial value problem as a system of
uncoupled (1+1)-dimensional partial differential equations, one for each spherical-harmonic $\ell$
mode. The numerical solution was augmented with analytical approximations for the early-time
\cite{Casals:2009xa} and late-time \cite{Casals:2012tb,Casals:Ottewill:2014} behavior of the RGF.

\subsection{Advantages}
\label{sec:advantages}

The RGF approach introduced in this paper has several distinct advantages. First, the regular part
of the self-field and the self-force are simply calculated by excluding the coincidence limit of
the Green function when computing the worldline integrals in (\ref{eq:Phisym1}) and
(\ref{eq:Fsym1}). This regularization procedure is valid for arbitrary worldlines, even accelerated
ones. This feature should be compared to the regularization procedures used in the more established
methods of mode-sum regularization and effective source. In both approaches the self-field and
self-force are regularized using parameters in the former and an effective source in
the latter. Both quantities must be derived and computed beforehand for the given worldline. In
addition, calculating the effective source can involve a significant fraction of the numerical
computing time.

Second, once the RGF has been computed for a given base point,
we can compute the regular parts of 
the self-field and self-force for \emph{any} worldline passing through that spacetime point, i.e.~all geodesics
as well as generic, accelerated worldlines.
This remarkable feature is unique compared to mode-sum and effective source methods, which can only compute the self-force for one worldline at a time. 
In this aspect, our method is complementary to other methods. The worldline integration method gives
the self-force for all worldlines but only at one base point, whereas mode-sum and effective-source 
methods compute the self-force at any point on a worldline but only for one worldline.

Third, our method admits geometrical interpretations and allows for quantitative comparisons between 
contributions from back-scattering and caustic echoes in determining the magnitude and sign of the 
self-field and self-force. Such a geometrical picture is lacking in other approaches to self-force computations. 
We anticipate this to carry through to gravitational self-force where one can make similar 
interpretations but only within the gauge choice made for the gravitational perturbations.

Fourth, our method allows for an offline/online decomposition of the problem. In the offline stage, we can 
devote as many computational resources as desired to produce a highly accurate numerical 
approximation of the RGF. The RGF can thus be computed accurately once and for all for a sufficient
number of field and base points. 
With the global RGF available from the offline stage, the online stage involves only cheap convolution integrals to compute the self-field and the self-force on 
a given worldline. The offline/online decomposition is particularly advantageous when computing the
RGF for a very narrow Gaussian where significant computational resources are necessary for computing
the RGF but not for evaluating worldline integrals. 

Fifth, and perhaps the most powerful advantage of our method, knowledge of the RGF \emph{and} of the regular self-force equations in terms 
of the RGF allows for the computation of higher order self-force effects with little further work. For 
example, once the scalar RGF is known one can simply perform the worldline integrals in the formal 
second order self-force expressions \cite{Galley:2010xn, Galley:2011te} to obtain the regular part of the 
second order self-force on an arbitrary worldline. In both mode-sum and effective source approaches the 
contributions at higher orders to the regularization parameters and effective source, respectively, need to 
be derived first, which is a nontrivial task, before nonlinear self-force computations can be performed. 

\subsection{Challenges}
\label{sec:challenges}

We mention four main challenges for the future applications of the worldline integration method. 

First, the computation of the RGF needs to be performed for each base point that a given worldline passes
through. In practice, one would require only a sufficiently dense distribution of base points for which the
symmetries of the background can be exploited. Nevertheless, the construction of such a distribution is
computationally expensive and will require large memory storage. Depending on how similar the solutions are from one base 
point to another, it is likely that the full space of solutions (parameterized by the base point) admits a 
reduced representation that is spanned by a compact set of judiciously selected solutions. Such a 
representation can be found with the Reduced Basis Method \cite{Field:2011mf}. Further reduced-order 
modeling techniques can be implemented to effectively predict the RGF associated with an arbitrary base 
point using a {\it surrogate model} \cite{Field:2013cfa} in place of solving for the full wave equation
separately for each base point. 
Surrogate models may provide a highly compressed and accurate representation of
the full space of approximate Green functions, which are also inexpensive to evaluate. 
These methods offer a promising avenue for future work, to solve the otherwise prohibitive computational and memory storage requirements involved with approximating the RGF at many base points. 

Second, obtaining increasingly accurate RGFs requires reducing the Gaussian width $\varepsilon$ and 
increasing the number of $\ell$ modes, which may become prohibitively expensive both in computing time
and memory storage. 
In Sec.~\ref{sec:gaussiandelta} we showed that decreasing $\varepsilon$ improves the self-field and self-force 
as $\varepsilon^{2}$. Likewise, an increase in the number of $\ell$ modes yields an improvement that 
scales as $1/\ell^{2}$. This improvement stems primarily from the fact that at early times (within the normal 
neighborhood) the direct $U(x,x') \delta(\sigma)$ part of the Hadamard form Green function, Eq. 
\eqref{eq:Hadamard}, is smeared out over the entire normal neighborhood by the finite sum over $\ell$ 
and contaminates the Green function in that region. A significant gain is therefore possible without 
decreasing $\varepsilon$ or increasing $\ell$, but by subtracting the $\ell$-decomposition of this direct 
part~\cite{Nolan-Capra2013,Casals:2012px} before summing over $\ell$.
Another approach is to adapt numerical methods for high-frequency wave propagation such as the 
``frozen Gaussian approximation'' \cite{LuYang} based on a paraxial approximation of the wave equation.
Yet another is to supplement the numerical approximation of the RGF with (semi-)analytical 
high-frequency/large-$\ell$ methods such as the geometrical optics approximation discussed in \cite{Zenginoglu:2012xe}, which was shown to capture the high-frequency behavior of the caustic echoes very accurately, and the large-$\ell$ asymptotics for the Green function multipolar
modes, which accurately capture the global four-fold singularity structure~\cite{Nolan-Capra2013,Casals:2012px}.
These methods may provide a practical alternative to the brute force approach of reducing the Gaussian
width.

Third, our method may not seem natural for computing self-consistent orbits as compared to, for example, 
the effective source method. The main challenge comes, again, from requiring the 
RGF at multiple base points. It should be possible to approximate the RGF at multiple base points for a
self-consistent evolution by either using a sufficiently dense distribution of base points along with 
interpolation, or using the reduced order modeling techniques \cite{Field:2011mf, Field:2013cfa} 
discussed above.

Fourth, extending the method to compute \emph{gravitational} self-force via the MiSaTaQuWa equation poses some technical challenges. The MiSaTaQuWa equation is formulated in terms of the derivatives of the Green function for a metric perturbation in Lorenz gauge. In Lorenz gauge, the metric perturbation on Schwarzschild spacetime may be decomposed in tensor spherical harmonics, leading to ten coupled 1+1D equations (or more precisely two sets, of 7 even and 3 odd-parity equations) \cite{Barack:2005nr}, which may be solved numerically with the methods described here. In principle, by starting with initial data in each component in turn, one may compute (an approximation to) the RGF and its derivatives. A naive method would require ten separate runs to compute the 100 components of the RGF, and computing derivatives of the RGF would require further runs with different initial data, as described in Sec.~\ref{sec:Kirchhoff}. An additional challenge is posed by the low multipoles $\ell \le 1$. These modes contain non-radiative physical content, related to the changes in mass and angular momentum of the system. In Lorenz gauge, initial-value formulations are also affected by linear-in-$t$ gauge mode instabilities in these modes, described in Sec.~V of Ref.~\cite{Dolan:2013roa}. A possible solution proposed there is to employ a generalized version of Lorenz gauge.

Unfortunately, in Lorenz gauge on Kerr spacetime the field equations cannot be separated into 1+1D form, and here we face a choice. Either we compute the RGF in Lorenz gauge by evolving multi-dimensional equations in the time domain (for example, the 2+1D approach of Ref.~\cite{Dolan:2013roa}), or we pursue a calculation in the radiation gauge \cite{Shah:2010bi, Keidl:2010pm, Shah:2012gu}. A key advantage of the latter approach is that the system is governed by \emph{ordinary} differential equations. On the other hand, the radiation gauge calculation is highly technical (relying on Hertz potentials and metric reconstruction through application of linear differential operators) and is formulated entirely within the frequency domain, which would seem to negate the advantages of the time-domain approach developed here. Reformulating the MiSaTaQuWa equation to make use of radiation-gauge Green functions would present an additional challenge. 

The resolution of these challenges should be the focus of future work to establish the worldline integration
method as a practical and accurate approach to the self-force problem.

\acknowledgments

A.C.O. and B.W. gratefully acknowledge support from Science Foundation Ireland under Grant
No.~10/RFP/PHY2847; B.W. also acknowledges support from the John Templeton Foundation New Frontiers
Program under Grant No.~37426 (University of Chicago) - FP050136-B (Cornell University).
C.R.G. was supported in part by NSF grants PHY-1316424, PHY-1068881, and CAREER grant PHY-0956189 
to the Caltech and by NASA grant NNX10AC69G. A.Z. was supported by NSF grant PHY-1068881 and by a 
Sherman Fairchild Foundation grant to Caltech. 
The authors additionally wish to acknowledge the SFI/HEA Irish Centre for High-End Computing
(ICHEC) for the provision of computational facilities and support (project ndast005b).

\bibliography{refs}

\end{document}